\documentclass[%
 aip,
 jmp,%
 amsmath,amssymb,
 reprint,%
]{revtex4-2}
\usepackage{hyperref}
\usepackage[all]{xy,xypic}
\usepackage{amsfonts,amssymb,amsmath,amsgen,amsopn,amsbsy,theorem,graphicx,epsfig}
\usepackage{eufrak,amscd,bezier,latexsym,mathrsfs,enumerate,multirow}
\usepackage[utf8]{inputenc}\usepackage[english]{babel}
\usepackage[dvipsnames]{xcolor}
\usepackage{academicons}
\usepackage{cases}
\usepackage[pagewise]{lineno}
\usepackage{graphicx}
\usepackage{dcolumn}
\usepackage{bm}

\begin{document}

\preprint{AIP/123-QED}

\title[]{The five dimensional transformed Weyl$-$Yang$-$Kaluza$-$Klein theory of gravity}

\author{H. KUYRUKCU}
\email{kuyrukcu@beun.edu.tr}
 \affiliation{Physics Department, Faculty of Science, Zonguldak B\"{u}lent Ecevit University,
 Zonguldak,
Turkey.}

\date{\today}

\begin{abstract}
We study the case in which the five dimensional theory is the transformed Weyl$-$Yang$-$Kaluza$-$Klein gravity. The dimensionally reduced equations of motion are derived by considering an alternative form of the main equation of the theory in the coordinate basis. The conformal transformation rules are applied to the invariants. We also discuss the possible specific cases and the new Lorentz force density term, in detail.
\end{abstract}

\maketitle

\section{Introduction}
Quadratic curvature Lagrangians have been used by many researchers to generalize or expand Einstein’s theory of general relativity for over a century (for early works, see, e.g.,\cite{Weyl1,Weyl2,Pauli,Weyl3,Eddington}). The simple alternative forms of the scalar curvature, $R$, could be the square version of the scalar curvature, $R^{2}$, the Ricci tensor, $R_{\mu\nu}R^{\mu\nu}$, or of the Riemann tensor, $R_{\mu\nu\lambda\sigma}R^{\mu\nu\lambda\sigma}$, and a linear combination of those, such as the well-known Gauss$-$Bonnet invariant\cite{Lanczos1,Lanczos2,Lanczos3}, as one step forward. Among them, mathematically, the most similar to the Yang$-$Mills gauge theory\cite{YangMills} is the following matter-free gravitational action quadratic equation in $R_{\mu\nu\lambda\sigma}$:
\begin{eqnarray}\label{action}
 I=\frac{1}{2 \kappa^{2}}\int_{\mathcal{M}}d^{4}x\,\sqrt{-g}\,{R}_{\mu\nu\lambda\sigma}{R}^{\mu\nu\lambda\sigma},
\end{eqnarray}
where ${\kappa}$ and $g$ represent the coupling constant and the determinant of $g_{\mu\nu}$, respectively, on a four-dimensional (4D) spacetime $\mathcal{M}$. By accepting that an affine connection is a Levi$-$Civita connection, i.e., torsion is absent, and considering a Palatini variational method, i.e., the metric $\{\mathfrak{g}\}$ and connection $\{\Gamma\}$ are assumed to be independent variables \cite{palatini,Misner:1973prb}, the connection variation $\delta\{\Gamma\}$ of this curvature-squared action (\ref{action}) provides the field equation without matter \cite{Lichnerowicz,Loos,Loos67,Yang}
\begin{eqnarray}\label{fe}
{D}_{\mu}{R}^{\mu}\,_{\nu\lambda\sigma}=0,
\end{eqnarray}
which is known as Yang’s gauge gravity equations in the literature. The third-order equations (\ref{fe}) are more complicated than well-established Einstein’s field equations; however, they include Einstein’s vacuum solutions in a natural manner. From the relation
\begin{eqnarray}\label{fe1e}
{D}_{\mu}{R}^{\mu}\,_{\nu\lambda\sigma}\equiv{D}_{\lambda}{R}_{\nu\sigma}-{D}_{\sigma}{R}_{\nu\lambda}=0,
\end{eqnarray}
which is derived from the second Bianchi identity $D_{[\rho}{R}_{\mu\nu]\lambda\sigma}=0$, where $D$ refers to the typical covariant derivative, the obvious vacuum solutions, ${R}_{\mu\nu}=0$ and ${R}_{\mu\nu}=\lambda g_{\mu\nu}$, for any constant $\lambda$, satisfy the equivalent equation (\ref{fe1e}). Conversely, the Palatini variation of the action (\ref{action}) with respect to the metric $\{\mathfrak{g}\}$ without matter can be written as follows \cite{stephenson,fairchild76,fairchild77}:
\begin{eqnarray}\label{fe2e}
{R}_{\mu\rho\lambda\sigma}R_{\nu}\,^{\rho\lambda\sigma}
-\frac{1}{4}g_{\mu\nu}{R}_{\tau\rho\lambda\sigma}R^{\tau\rho\lambda\sigma}=0.
\end{eqnarray}
The field equations (\ref{fe2e}) are initially considered to eliminate nonphysical solutions\cite{fairchild76,pavelle2,pavelle3} even though numerous physical ones exist in the literature\cite{pavelle1,thompson,thompson1,ni,pavelle4,baskal,kuyrukcu2013,kuyrukcu2021} (for more historical notes, see, e.g.,\cite{kuyrukcu2021,dean}). Besides, Baekler and Yasskin \cite{Baekler} noted that the standard variation of (\ref{action}) produces the equations, which is fourth order for a metric
\begin{eqnarray}\label{fe3e}
{R}_{\mu\rho\lambda\sigma}R_{\nu}\,^{\rho\lambda\sigma}
-\frac{1}{4}g_{\mu\nu}{R}_{\tau\rho\lambda\sigma}R^{\tau\rho\lambda\sigma}
+2D^{\lambda}D^{\sigma}{R}_{\sigma\mu\lambda\nu}=0,
\end{eqnarray}
which is Eddington's equation \cite{Eddington}. They also noted that “if a metric (or tetrad) satisfies the torsion-free vacuum equations (\ref{fe}) and (\ref{fe2e}), it also satisfies the vacuum Eddington equation (\ref{fe3e}), but not vice versa”. For the case in which torsion is zero, but the matter Lagrangian term exists and ${\kappa}=1$, the field equations (\ref{fe}) and (\ref{fe2e}) become
\begin{eqnarray}\label{fe0e}
&&{D}_{\mu}{R}^{\mu}\,_{\nu\lambda\sigma}=S_{\lambda\sigma\nu},\\
&&{R}_{\mu\rho\lambda\sigma}R_{\nu}\,^{\rho\lambda\sigma}
-\frac{1}{4}g_{\mu\nu}{R}_{\tau\rho\lambda\sigma}R^{\tau\rho\lambda\sigma}=T_{\mu\nu},\label{fe4e}
\end{eqnarray}
where $S_{\lambda\sigma\nu}$ represents the canonical spin tensor and $T_{\mu\nu}$ represents the canonical energy-momentum tensor in the aspect of the quadratic Poincar\'{e} gauge theory of gravity\cite{Baekler}. Camenzind and Fairchild also considered the equation (\ref{fe4e}) as an energy-momentum tensor of this alternative theory rather than the typical field equations \cite{fairchild77,cam75}. Kilmister et al. \cite{Kilmister} also introduced the current term $S_{\lambda\sigma\nu}$ , which satisfies a covariant conservation property of the form $D_{\nu}S^{\lambda\sigma\nu}=0$, but they could not clearly define what it was. Besides, the cyclic symmetry property of  $S_{\lambda\sigma\nu}$ , i.e., $S_{[\lambda\sigma\nu]}=0$ , is proposed by \"{O}ktem \cite{oktem}. Furthermore, by considering Einstein’s field equations, $G_{\mu\nu}=T_{\mu\nu}$,  along with the field equation (\ref{fe1e}), the source term can be written in the following forms
\begin{eqnarray}\label{fe5e}
S_{\lambda\sigma\nu}=D_{\lambda}\left(T_{\nu\sigma}-\frac{1}{2}g_{\nu\sigma}T\right)
-D_{\sigma}\left(T_{\nu\lambda}-\frac{1}{2}g_{\nu\lambda}T\right),
\end{eqnarray}
where $T_{\nu\sigma}$ represents the covariantly conservative, $D_{\mu}T^{\mu\nu}=0$, energy-momentum tensor, whose trace is $T=T_{\nu}\,^{\nu}$. This current density term $S_{\lambda\sigma\nu}$ in (\ref{fe5e}) was first used for Yang$–$Mills field equations with the SO(3,1) gauge group by Camenzind\cite{cam75,cam75a,cam77,cam78,cam78a}. The metric is a nondynamic variable, i.e., a priori in this sense. Later, Cook used Camenzind’s matter current term (\ref{fe5e}) by considering a formal analogy between Einstein’s theory of relativity and classical electrodynamics, which means the connection $\{\Gamma\}$ and Riemann tensor $\{R\}$ correspond to the vector potential $\{A\}$ and electromagnetic field tensor $\{F\}$, respectively, to solve vacuum energy, cosmological constant, and dark energy problems\cite{Cook} (see also, e.g., Chen et al. \cite{Chen}). However, even if the term (\ref{fe5e}) is suggested to solve the source-term problem of this simple gravity model, it cannot be obtained from an action principle, meaning that it is inconsistent.

The main motivation in this work is to extend and generalize our previous results\cite{halil13} to the Einstein frame in which the Ricci scalar has its canonical form by considering that the five-dimensional (5D) theory is the Weyl$-$Yang$-$Kaluza$-$Klein (WYKK) theory of gravity (for non-Abelian WYKK theory, see also \cite{halil14}). Weyl was the first to consider using the $R_{\mu\nu\lambda\sigma}R^{\mu\nu\lambda\sigma}$ invariant in the action to unify gravitation with electromagnetism by employing the principle of gauge invariance, the vacuum gravitational field equation (\ref{fe}) was proposed by Yang in an integral formalism for gauge fields without additional equations (\ref{fe2e}) and (\ref{fe3e}), and Kaluza and Klein (KK) assumed that gravitation and electromagnetism can be unified in the 5D spacetime \cite{Kaluza,Klein,Mandel,KleinN}. Then, we can prefer to call this higher-dimensional model the WYKK theory of modified gravity rather than just KK reduction of a quadratic gravity even though they are all different situations. Conversely, we can note that the field equations (\ref{fe}), (\ref{fe2e}), (\ref{fe0e}), (\ref{fe4e}), and (\ref{fe5e}) are also known as Stephenson$-$Kilmister$-$Yang$-$Camenzind equations in the literature. In this sense, we take advantage of the horizontal lift basis that is given by Misner et al. \cite{Misner:1973prb} for faster calculations rather than the differential forms available in the literature \cite{Pope,Perry}. Moreover, the dimensionally reduced vielbein components of the curvature tensor, which are important for the considered model, are explicitly presented in the opened-notation form rather than the compact-notation form. The possible special cases are also investigated and discussed, along with the new Lorentz force density term, in detail. Before we attempt to obtain the reduced conformal equations, we wish to revisit our previous work \cite{halil13} to explain how we can use an alternative equation (\ref{fe1e}) to derive the field equations for the reader’s convenience.

The layout of this paper is as follows: In Section \ref{sec2}, we present a brief review of the 5D WYKK theory of gravity and derive new accurate field equations, in particular. In Section \ref{sec3}, the transformed equations of motion are found for the generalized KK metric ansatz. Section \ref{sec4} is concerned with obtaining the reduced form of the Kretschmann invariant\cite{Kretschmann}, $R_{\mu\nu\lambda\sigma}R^{\mu\nu\lambda\sigma}$,  both by applying a circle reduction mechanism and performing the conformal scale transformations in the so-called Einstein frame. Section \ref{sec5} is devoted to investigating special cases of the reduced equations. In Section \ref{sec6}, we finally present our findings and future work.

\section{\label{sec2}Revisit of the Reduced Equations of the WYKK Theory}

As usual, the 5D standard KK action, $\hat{I}$, is given by
\begin{eqnarray}\label{kkaction}
\hat{I}=\frac{1}{2\hat\kappa^{2}}\int_{\hat{\mathcal{M}}}d^{4}x\,dy\,\sqrt{-\hat{g}}\,\hat R,
\end{eqnarray}
on the 5D manifold $\hat{\mathcal{M}}$ with compactification of the type $\hat{\mathcal{M}}=\mathcal{M}\times\mathcal{S}^{1}$, where $\mathcal{M}$ is the usual 4D spacetime as used previously, and $\mathcal{S}^{1}$ has the geometry of a small circle. Conversely, $\hat R$, $\hat{\kappa}$, and $\hat g$ refer to the ordinary 5D Ricci curvature; the coupling constant, which is related to the 4D coupling constant, $\kappa$ and  the radius of the fifth dimension, $r$, as ${\hat\kappa^{2}}={\kappa^{2}}\int dy=2\pi r{\kappa^{2}}$; and the determinant of  $ \hat g_{MN}$, i.e., $\hat g=det(\hat g_{MN})$, respectively. Here the classical metric ansatz of the theory, which is independent of the fifth dimension, becomes
\begin{eqnarray}\label{kkmetric}
\hat g_{MN}(x,y)=\left(\begin{array}{r|r}
g_{\mu\nu}(x)+\varphi^{2}(x) A_{\mu}(x)A_{\nu}(x)&\varphi^{2}(x) A_{\mu}(x)
\\[0.3em] \hline  \rule{0pt}{1.0\normalbaselineskip}
\varphi^{2}(x) A_{\nu}(x)&\varphi^{2}(x)
\end{array}\right),
\end{eqnarray}
and hence $\sqrt{-\hat g}=\varphi\sqrt{-det( g_{\mu\nu})}=\varphi\sqrt{-g}$. Moreover,
$\varphi(x)$ is the dilaton field, $A_{\mu}(x)$ is the gauge potential, and $g_{\mu\nu}(x)$ is the 4D metric tensor in the usual manner. Kaluza also assumed that the derivatives of all the fields with respect to the new fifth coordinate, $y$, vanish \cite{Kaluza}. In what follows, the hatted$/$unhatted fields are 5D$/$4D. We can also introduce 5D coordinates as $\hat x^A=(x^\mu,x^5)$ or, in short, $\hat x=(x,y)$, where the capital curved Latin indices are~$A,B,...=0, 1, 2, 3, 5$, and the Greek curved indices are $\mu,\nu,...=0, 1, 2, 3$.

For shortcut calculations, we can consider the Maurer$-$Cartan exterior forms rather than the coordinate basis as a type of strategy. In this fashion, the 5D line element, $\hat G(x,y)=\hat g_{ MN}dx^{M}dx^{N}$, takes the form
\begin{eqnarray}
\hat G(x,y)=\hat \eta_{\hat A\hat B}\hat E^{\hat A}\otimes\hat E^{\hat B},
\end{eqnarray}
where the coframes one-form  become
\begin{eqnarray}\label{coframes1}
\hat E^{\hat \mu}(x,y)= E^{\hat \mu}(x),\qquad\qquad\qquad\qquad\hat E^{\hat 5}(x,y)= \varphi(x)\left[A(x)+dy\right],
\end{eqnarray}
and the 5D  Minkowski or orthonormal metric is $\hat\eta_{\hat A\hat B}=diag(\eta_{\hat\mu\hat\nu},+1)$. Hence, we can write the 4D line element as $G(x)=\eta_{\hat\mu\hat\nu}E^{\hat \mu}(x)\otimes E^{\hat \nu}(x)$ with $\eta_{\hat\mu\hat\nu}=diag(-1,+1,+1,+1)$, and  the gauge fields as $A(x)=A_{\hat\mu}(x)E^{\hat \mu}(x)$, where all hatted indices refer to the orthonormal basis, i.e., the flat indices. We can now obtain the dimensionally reduced components of the spin connection one-form, $\hat\Gamma_{\hat A\hat B}$, and the curvature two-form $\hat{\mathcal{R}}^{\hat A}\,_{\hat B}$ by considering Maurer$-$Cartan structure equations (see, e.g., \cite{dereli90,kuyrukcu2013}). However, we prefer to transform the desired components in opened-notation form rather than the compact-notation form to better understand the structure of these equations as follows:
\begin{eqnarray}\label{anholomoniccristofel}
&&\hat\Gamma^{\hat\mu}\,_{\hat\nu\hat\lambda}=\Gamma^{\hat\mu}\,_{\hat\nu\hat\lambda},
\qquad\qquad\quad
\hat\Gamma^{\hat5}\,_{\hat\mu\hat\nu}= \frac{1}{2} \varphi F_{\hat\mu\hat\nu},
\qquad\qquad\quad
\hat\Gamma^{\hat\mu}\,_{\hat5\hat\nu}
=\hat\Gamma^{\hat\mu}\,_{\hat\nu\hat5}=-\frac{1}{2} \varphi F^{\hat\mu}\,_{\hat\nu},
\nonumber\\&&
\hat\Gamma^{\hat\mu}\,_{\hat5\hat5}=-\varphi^{-1}\varphi^{\hat\mu},
\qquad\quad\quad
\hat\Gamma^{\hat5}\,_{\hat\mu\hat5}=\varphi^{-1}\varphi_{\hat\mu},
\end{eqnarray}
and
\begin{eqnarray}\label{riem}
&&\hat{R}^{\hat\mu}\,_{\hat\nu\hat\lambda\hat\sigma}=R^{\hat\mu}\,_{\hat\nu\hat\lambda\hat\sigma}
-\frac{1}{4}\varphi^{2}\left(2 F^{\hat\mu}\,_{\hat\nu}F_{\hat\lambda\hat\sigma}+F^{\hat\mu}\,_{\hat\lambda}F_{\hat\nu\hat\sigma}
-F^{\hat\mu}\,_{\hat\sigma}F_{\hat\nu\hat\lambda}\right), \nonumber\\
&&\hat{R}^{\hat\mu}\,_{\hat5\hat\lambda\hat\sigma}=-\frac{1}{2}\varphi D^{\hat\mu}F_{\hat\lambda\hat\sigma}
-\frac{1}{2}\left( 2\varphi^{\hat\mu} F_{\hat\lambda\hat\sigma}
+\varphi_{\hat\lambda} F^{\hat\mu}\,_{\hat\sigma}-\varphi_{\hat\sigma}F^{\hat\mu}\,_{\hat\lambda}\right),\nonumber\\
&&\hat{R}^{\hat\mu}\,_{\hat5\hat\lambda\hat5}=-\varphi^{-1}D_{\hat\lambda}\varphi^{\hat\mu}
-\frac{1}{4}\varphi^{2}F^{\hat\mu\hat\rho}F_{\hat\rho\hat\lambda}.
\end{eqnarray}
Note that $\hat{R}^{\hat\lambda}\,_{\hat\sigma\hat5\hat\mu}=-\hat{R}^{\hat\mu}\,_{\hat5\hat\lambda\hat\sigma}$, and $D_{\hat\mu}\varphi_{\hat\nu}=D_{\hat\nu}\varphi_{\hat\mu}$ as well as $D$, $F_{\hat\mu\hat\nu}$, and $\varphi_{\hat\mu}$ refer to the 4D covariant derivative, the electromagnetic field tensor i.e., $F_{\hat\mu\hat\nu}=\partial_{\hat\mu} A_{\hat\nu}-\partial_{\hat\nu} A_{\hat\mu}$, and  $\varphi_{\hat\mu}=D_{\hat\mu}\varphi=\partial_{\hat\mu}\varphi$. The same results, (\ref{anholomoniccristofel}) and (\ref{riem}), have already been discussed in the language of horizontal lift basis (see, e.g.,\cite{halil13}, as well as\cite{Lee83} for  metric signature $(+,-,-,-,-)$, and  $\varphi(x)=1$ ). Conversely, for later convenience, we set
\begin{eqnarray}\label{ortricci}
{\hat R}_{\hat\mu\hat\nu}\equiv\mathcal{\overline{P}}_{\hat\mu\hat\nu},\qquad\qquad\qquad
{{\hat R}}_{\hat5\hat\nu}\equiv\mathcal{\overline{Q}}_{\hat\nu},\qquad\qquad\qquad
\hat R_{\hat5\hat5}\equiv\mathcal{\overline{U}},
\end{eqnarray}
and in terms of vielbeins, which are the orthonormal basis vectors, the 5D metric takes the form $\hat g_{ MN}=\hat h^{\hat A}\,_{M}\hat h^{\hat B}\,_{N}\hat\eta_{\hat A\hat B}$, where the vielbeins and the inverse vielbeins are
\begin{eqnarray}\label{viel}
\hat h^{\hat A}\,_{M} =\left(
\begin{array}{cc}
\hat h^{\hat\mu}\,_{\nu}   &\hat h^{\hat\mu}\,_{5}  \\
 \hat h^{\hat5}\,_{\nu}& \hat h^{\hat5}\,_{5}
\end{array}
\right)=\left(
\begin{array}{cc}
 h^{\hat\mu}\,_{\nu}   & 0 \\
\varphi A_{\nu} & \varphi
\end{array}
\right),
\quad
\hat h^{M}\,_{\hat A}=\left(
\begin{array}{cc}
 \hat h^{\nu}\,_{\hat\mu}   &\hat h^{\nu}\,_{\hat5}  \\
 \hat h^{5}\,_{\hat\mu}& \hat h^{5}\,_{\hat5}
\end{array}
\right)=\left(
\begin{array}{cc}
 h^\nu\,_{\hat\mu}   & 0 \\
- A_{\hat\mu} & \varphi^{-1}
\end{array}
\right),
\end{eqnarray}
which satisfy $\hat h^{\hat A}\,_{M}\hat h^{M}\,_{\hat B}=\hat \delta^{\hat A}\,_{\hat B}$ ($h^{\hat \mu}\,_{\rho} h^{\rho}\,_{\hat\nu}=\delta^{\hat\mu}\,_{\hat\nu}$) and $\hat h^{M}\,_{\hat B}\hat h^{\hat B}\,_{N}=\hat \delta^{M}\,_{N}$  ($ h^{\mu}\,_{\hat\rho} h^{\hat\rho}\,_{\nu}=\delta^{\mu}\,_{\nu}$) in the five (four) dimensions. Hence, the final results of the components of the 5D Ricci tensor are calculated by considering the vielbeins (\ref{viel}) in the coordinate basis as follows (for the metric signature $(+,-,-,-,-)$, see, e.g.,\cite{Liu:1997fg}):
\begin{eqnarray}\label{}
&&{\hat R}_{\mu\nu}=\mathcal{\overline{P}}_{\mu\nu}+\varphi A_{\mu}\mathcal{\overline{Q}}_{\nu}
+\varphi A_{\nu}\mathcal{\overline{Q}}_{\mu}+\varphi^{2}A_{\mu}A_{\nu}\mathcal{\overline{U}}
 ,\nonumber\\
&&{{\hat R}}_{5\nu}=\varphi\mathcal{\overline{Q}}_{\nu}+\varphi^{2}A_{\nu}\mathcal{\overline{U}}
 ,\nonumber\\
&&\hat R_{55}=\varphi^{2}\mathcal{\overline{U}},
\end{eqnarray}
where
\begin{eqnarray}\label{ricciten}
&&\mathcal{\overline{P}}_{\mu\nu}
=R_{\mu\nu}-\frac{1}{2}\varphi^{2}
F_{\mu\rho}F_{\nu}\,^{\rho}-\varphi^{-1}D_{\mu}\varphi_{\nu} ,\nonumber\\
&&{\overline{Q}}_{\nu}
=-\frac{1}{2}\varphi\left({D}_{\rho}F^{\rho}\,_{\nu}
+3\varphi^{-1}\varphi_{\rho}F^{\rho}\,_{\nu}\right) ,\nonumber\\
&&\mathcal{\overline{U}}
=-\varphi^{-1}\left(D_{\rho}\varphi^{\rho}-\frac{1}{4}\varphi^{3} F_{\rho\tau}F^{\rho\tau}
\right).
\end{eqnarray}
The next step is to obtain the corresponding equations of motion. Thus, we consider the 5D Einstein equations $\hat G_{AB}=0$ or, equivalently, $\hat R_{AB}=0$; i.e., the Ricci curvature tensor vanishes, as usual. Hence, we have
\begin{numcases} {\hat R_{AB}=0,\quad\Rightarrow\quad}
\hat R_{55}=0,& $\quad\Rightarrow\quad \mathcal{\overline{U}}=0,$\nonumber\\
{{\hat R}}_{5\nu}=0,\quad $and$ \quad\mathcal{\overline{U}}=0, &$\quad\Rightarrow\quad  \mathcal{\overline{Q}}_{\nu}=0,$\nonumber\\
{\hat R}_{\mu\nu}=0,\quad $and$ \quad\mathcal{\overline{U}}=0,\quad$and$ \quad \mathcal{\overline{Q}}_{\nu}=0,& $\quad\Rightarrow\quad \mathcal{\overline{P}}_{\mu\nu}=0,$
\end{numcases}
which means that the solution set, $S_{KK}$, of the standard KK theory turns out to be
\begin{eqnarray}\label{solutionKK}
S_{KK}=\Big\{\mathcal{\overline{P}}_{\mu\nu}=0,\;~\mathcal{\overline{Q}}_{\nu}=0,\;~\mathcal{\overline{U}}=0\Big\}.
\end{eqnarray}
Let us now investigate the dimensionally reduced field equations of the WYKK theory, where the source-free field equations become $\hat{D}_{ A}\hat{R}^{ A}\,_{BCD}=0$ in the five dimensions. It is useful to consider the equivalent form (\ref{fe1e}) of field equations (\ref{fe}) to obtain the computational advantage in the orthonormal chart, as follows:
\begin{eqnarray}\label{bianc2}
\hat{D}_{\hat A}\hat{R}^{\hat A}\,_{\hat B\hat C\hat D}\equiv\hat{ D}_{\hat C}\hat{R}_{\hat B\hat D}-\hat{ D}_{\hat D}\hat{ R}_{\hat B\hat C}.
\end{eqnarray}
If we define
\begin{eqnarray}
\hat{D}_{\hat A}\hat{R}^{\hat A}\,_{\hat\nu\hat\lambda\hat\sigma}\equiv\mathbb{{P}_{\hat\nu\hat\lambda\hat\sigma}},\quad
\hat{D}_{\hat A}\hat{R}^{\hat A}\,_{\hat5\hat\lambda\hat\sigma}\equiv\mathbb{S_{\hat\lambda\hat\sigma}},\quad
\hat{D}_{\hat A}\hat{R}^{\hat A}\,_{\hat\nu\hat\lambda\hat5}\equiv\mathbb{Q_{\hat\nu\hat\lambda}},\quad
\hat{D}_{\hat A}\hat{R}^{\hat A}\,_{\hat5\hat\lambda \hat5}\equiv\mathbb{U_{\hat\lambda}},
\end{eqnarray}
and by substituting the expressions (\ref{ortricci}) into the equation (\ref{bianc2}) together with the connection terms (\ref{anholomoniccristofel}), performing some manipulations,
and then employing the vielbeins (\ref{viel})   to convert the resulting equations into the desired field equations  that are written in coordinate basis, we find reduced equations in the following forms:
\begin{eqnarray}
&&\hat{D}_{A}\hat{R}^{A}\,_{\nu\lambda\sigma}=\mathbb{{P}_{\nu\lambda\sigma}}
+\varphi A_{\nu}\mathbb{S_{\lambda\sigma}}
+\varphi A_{\sigma}\mathbb{Q_{\nu\lambda}}
-\varphi A_{\lambda}\mathbb{Q_{\nu\sigma}}
+\varphi^{2}A_{\nu}A_{\sigma}\mathbb{U_{\lambda}}
-\varphi^{2}A_{\nu}A_{\lambda}\mathbb{U_{\sigma}},\nonumber\\
&&\hat{D}_{A}\hat{R}^{A}\,_{5\lambda\sigma}=\varphi\mathbb{S_{\lambda\sigma}}
+\varphi^{2}A_{\sigma}\mathbb{U_{\lambda}}
-\varphi^{2}A_{\lambda}\mathbb{U_{\sigma}},\nonumber\\
&&\hat{D}_{A}\hat{R}^{A}\,_{\nu\lambda5}=-\hat{D}_{A}\hat{R}^{A}\,_{\nu5\lambda}=\varphi\mathbb{Q_{\nu\lambda}}
+\varphi^{2}A_{\nu}\mathbb{U_{\lambda}},\nonumber\\
&&\hat{D}_{A}\hat{R}^{A}\,_{5\lambda 5}=-\hat{D}_{A}\hat{R}^{A}\,_{55\lambda}=\varphi^{2}\mathbb{U_{\lambda}},
\end{eqnarray}
where
\begin{eqnarray}
 & &{\mathbb{P_{\nu\lambda\sigma}}}= D_{\lambda}\mathcal{\overline{P}}_{\nu \sigma}-D_{\sigma}\mathcal{\overline{P}}_{\nu \lambda}
+ \frac{1}{2}\varphi \left( F_{\nu \sigma}\mathcal{\overline{Q}}_{\lambda}-F_{\nu \lambda}\mathcal{\overline{Q}}_{\sigma}
+ 2 F_{\lambda\sigma}\mathcal{\overline{Q}}_{\nu} \right),\label{p1}  \\
& & \mathbb{S_{\lambda\sigma}}=D_{\lambda}\mathcal{\overline{Q}}_{\sigma}-D_{\sigma}\mathcal{\overline{Q}}_{\lambda}- \frac{1}{2}\varphi \left(F_{\lambda}\,^{\rho}\mathcal{\overline{P}}_{\rho\sigma}-F_{\sigma}\,^{\rho}
\mathcal{\overline{P}}_{\rho\lambda}\right)
+\varphi F_{\lambda\sigma}\mathcal{\overline{U}},\label{p2}   \\
 & & \mathbb{Q_{\nu\lambda}}=D_{\lambda}\mathcal{\overline{Q}}_{\nu} +\varphi^{-1}( \varphi_{\nu} \mathcal{\overline{Q}}_{\lambda} +  \varphi_{\lambda}\mathcal{\overline{Q}}_{\nu})
 +\frac{1}{2}\varphi   F_{\nu}\,^{\rho} \mathcal{\overline{P}}_{\rho\lambda}
 -\frac{1}{2}\varphi F_{\nu\lambda}\mathcal{\overline{U}}, \label{p3}  \\
 & & \mathbb{U_{\lambda}}=D_{\lambda}\mathcal{\overline{U}}
 +\varphi^{-1}\varphi_{\lambda}\mathcal{\overline{U}}
-\frac{1}{2} \varphi F_{\lambda}\,^{\rho}\mathcal{\overline{Q}}_{\rho}
- \varphi^{-1}\varphi^{\rho} \mathcal{\overline{P}}_{\rho\lambda}.\label{p4}
\end{eqnarray}
There is no doubt that equations (\ref{p1})$-$(\ref{p4}) naturally contain  patterns of  (\ref{ricciten}), as expected. As a result, we have
\begin{numcases} {\hat{D}_{A}\hat{R}^{A}\,_{BCD}=0,\;~\Rightarrow\;~}
\hat{D}_{A}\hat{R}^{A}\,_{5\lambda 5}=0, &$\;~\Rightarrow\;~ \mathbb{U_{\lambda}}=0,$\nonumber\\
\hat{D}_{A}\hat{R}^{A}\,_{\nu\lambda5}=0,\;~  $and$ \;\mathbb{U_{\lambda}}=0, &$ \;~\Rightarrow\;~\mathbb{Q_{\nu\lambda}}=0,$\nonumber\\
\hat{D}_{A}\hat{R}^{A}\,_{5\lambda\sigma}=0,\;~ $and$ \;\mathbb{U_{\lambda}}=0, &$\;~\Rightarrow\;~ \mathbb{S_{\lambda\sigma}}=0,$
\nonumber\\
\hat{D}_{A}\hat{R}^{A}\,_{\nu\lambda\sigma}=0,\;~ $and$ \; \mathbb{U_{\lambda}}=0,\nonumber\\ $and$ \;~ \mathbb{Q_{\nu\lambda}}=0,\;~$and$ \; \mathbb{S_{\lambda\sigma}}=0, &$\;~\Rightarrow\;~ \mathbb{{P}_{\nu\lambda\sigma}}=0.$
\end{numcases}
In this case, the set of solution of the WYKK theory obviously becomes
\begin{eqnarray}
S_{WYKK}=\Big\{\mathbb{{P}_{\nu\lambda\sigma}}=0,\;~\mathbb{S_{\lambda\sigma}}=0,\;~\mathbb{Q_{\nu\lambda}}=0,\; \mathbb{U_{\lambda}}=0  \Big\}.
\end{eqnarray}
As can be easily seen,  equation (\ref{p2}) can be obtained from (\ref{p3})   using  $\mathbb{S_{\lambda\sigma}}=\mathbb{Q_{\sigma\lambda}}-\mathbb{Q_{\lambda\sigma}}$, meaning that  equation (\ref{p2}) is not necessary. Hence, the final forms of dimensionally reduced field equations are given by
\begin{eqnarray}
 & & D_{\lambda}\mathcal{\overline{P}}_{\nu \sigma}-D_{\sigma}\mathcal{\overline{P}}_{\nu \lambda}
+ \frac{1}{2}\varphi \left( F_{\nu \sigma}\mathcal{\overline{Q}}_{\lambda}-F_{\nu \lambda}\mathcal{\overline{Q}}_{\sigma}
+ 2 F_{\lambda\sigma}\mathcal{\overline{Q}}_{\nu} \right)=0,\label{proper1}    \\
 & & D_{\nu}\mathcal{\overline{Q}}_{\lambda} +\varphi^{-1}( \varphi_{\nu} \mathcal{\overline{Q}}_{\lambda} +  \varphi_{\lambda}\mathcal{\overline{Q}}_{\nu})
 +\frac{1}{2}\varphi   F_{\nu}\,^{\rho} \mathcal{\overline{P}}_{\rho\lambda}
 -\frac{1}{2}\varphi F_{\nu\lambda}\mathcal{\overline{U}}=0,\label{proper2}  \\
 & & D_{\lambda}\mathcal{\overline{U}}
 +\varphi^{-1}\varphi_{\lambda}\mathcal{\overline{U}}
-\frac{1}{2} \varphi F_{\lambda}\,^{\rho}\mathcal{\overline{Q}}_{\rho}
- \varphi^{-1}\varphi^{\rho} \mathcal{\overline{P}}_{\rho\lambda}=0,\label{proper3}
\end{eqnarray}
and any solutions of the KK theory, (\ref{solutionKK}), solve the reduced equations of the WYKK theory (\ref{proper1})$–$(\ref{proper3}). It is essential here to note that the $\left\{\mathcal{\overline{P}}_{\mu\nu},\mathcal{\overline{Q}}_{\nu},\mathcal{\overline{U}}\right\}$ set (\ref{ricciten}) can turn into the $\left\{\mathcal{{P}}_{\mu\nu},\mathcal{{Q}}_{\nu},\mathcal{{U}}\right\}$ set
via
\begin{eqnarray}\label{}
\mathcal{\overline{P}}_{\mu\nu}=\mathcal{P}_{\mu\nu}
 ,\qquad\qquad\qquad
\mathcal{\overline{Q}}_{\nu}=-\frac{1}{2}\varphi\mathcal{Q}_{\nu}
 ,\qquad\qquad\qquad
\mathcal{\overline{U}}=-\varphi^{-1}\mathcal{U}.
\end{eqnarray}
The reduced field equations, which can be written by considering the $\left\{\mathcal{{P}}_{\mu\nu},\mathcal{{Q}}_{\nu},\mathcal{{U}}\right\}$ set, have already  been given in our previous work\cite{halil13}. However, the expressions (\ref{proper1})$-$(\ref{proper3}) containing the $\left\{\mathcal{\overline{P}}_{\mu\nu},\mathcal{\overline Q}_{\nu},\mathcal{\overline{U}}\right\}$ set clearly seem to be  more accurate than those containing the $\left\{\mathcal{{P}}_{\mu\nu},\mathcal{{Q}}_{\nu},\mathcal{{U}}\right\}$~set, and we now have the three equations corresponding to the three variables. Another way of obtaining the field equations is the 5D WYKK action, which is given by
\begin{eqnarray}\label{acti}
\hat{I}=\frac{1}{2\hat\kappa^{2}}\int_{\hat{\mathcal{M}}}d^{4}x\,dy\,\sqrt{-\hat{g}}\,\hat R_{ABCD}\hat R^{ABCD}.
\end{eqnarray}
By substituting the dimensionally reduced form of the
Kretschmann invariant, $\mathcal{\hat {K}}=\hat R_{ABCD}\hat R^{ABCD}$, which is
found to be in \cite{halil13}, into action (\ref{acti}) and  dropping the total derivative terms that can come out with the help of Leibniz rule from the action gives
\begin{eqnarray}\label{inv1}
&&I=\frac{1}{2\kappa^{2}}\int_{{\mathcal{M}}}d^{4}x\,\sqrt{-{g}}\varphi \,\Big[ R_{\mu\nu\lambda\sigma}R^{\mu\nu\lambda\sigma}-\frac{3}{2}\varphi^{2} R_{\mu\nu\lambda\sigma}F^{\mu\nu}F^{\lambda\sigma}
+\frac{3}{8}\varphi^{4}F_{\mu\nu}F^{\mu\nu}F_{\lambda\sigma}F^{\lambda\sigma}
\nonumber\\&&\quad
+\frac{5}{8}\varphi^{4}F_{\mu\nu}F^{\nu\lambda}F_{\lambda\sigma}F^{\sigma\mu}
+6\left(\varphi_{\lambda}\varphi^{\lambda} F_{\mu\nu}F^{\mu\nu}
-\varphi_{\mu}\varphi^{\lambda}F_{\lambda\nu}F^{\nu\mu}\right)
\nonumber\\&&\quad
+ 4\varphi\left( \varphi^{\lambda} F^{\mu\nu}+\varphi^{\mu} F^{\lambda\nu} \right)D_{\lambda}F_{\mu\nu}
-\varphi^{2}F^{\mu\nu}D_{\lambda}D^{\lambda}F_{\mu\nu}
-2 \varphi  F^{\mu\lambda}F^{\nu}\,_{\lambda}D_{\mu}\varphi_{\nu}
\nonumber\\&&\quad
-4\varphi^{-2}\varphi^{\nu}D^{\mu}D_{\mu}\varphi_{\nu}\Big].
\end{eqnarray}
Finally, the independent variation of action, $I[\mathfrak{g},\Gamma,F,\varphi_{\mu}]$, (\ref{inv1}) with respect to the four variables gives the four field equations. The Palatini variation of the above action with respect to the metric $\{\mathfrak{g}\}$ should directly produce the dimensionally reduced equations of (\ref{fe2e}). However, we cannot directly obtain equations (\ref{proper1})$–$(\ref{proper3}) from the variational principle, as expected. For instance, the connection variation $\delta\{\Gamma\}$ leads to an equation that can transform into (\ref{proper1}) by using various identities (which are given by Başkal and Kuyrukcu\cite{halil13}). Additionally, to obtain expressions that can be written in forms of (\ref{proper2}) and (\ref{proper3}), we should vary the action with respect to the field tensor $\{F\}$  and the partial derivative of the boson fields $\{\varphi_{\mu}\}$ rather than the usual variables $\{A\}$ and $\{\varphi\}$, respectively; otherwise we cannot obtain the proper equations \cite{celik}.

\section{The Reduced Field Equations from the Transformed WYKK theory}
\label{sec3}
To obtain the field equations that are not only  dimensionally reduced but also transformed from the conformal rescaling procedure, it is useful to first write the D-dimensional Weyl-rescaled Ricci scalar, $\widetilde{R}$,\cite{Hawking1999,Dabrowski:2008kx} as
\begin{eqnarray}
\widetilde{R}=\xi^{-2}\Big[R-2(D-1)\xi^{-1}D_{\mu}\xi^{\mu}-(D-1)(D-4)\xi^{-2}\xi_{\mu}\xi^{\mu}\Big],
\end{eqnarray}
where the tilde quantities denote fields in the Einstein frame, as usual, under the Weyl rescaling of the metric with a conformal factor, $\xi(x)$, as follows:
\begin{eqnarray}
\widetilde{g}_{\mu\nu}(x)=\xi^{2}(x)g_{\mu\nu}(x).
\end{eqnarray}
After that, for $D=5$, the KK action (\ref{kkaction}) changes
\begin{eqnarray}
\widetilde{\hat I}&=&\frac{1}{2\hat\kappa^{2}}\int_{\hat{\mathcal{M}}}d^{4}x\,dy\,\sqrt{-\widetilde{\hat g}}\,\widetilde{\hat R},\label{invari1}\\
&=&\frac{1}{2\hat\kappa^{2}}\int_{\hat{\mathcal{M}}}d^{4}x\,dy\,\sqrt{-g}\,\xi^{5}\varphi\xi^{-2}
\left[\hat R-8\xi^{-1}\hat D_{A}\xi^{A}-4\xi^{-2}\xi_{A}\xi^{A}\right],
\end{eqnarray}
by using $\sqrt{-\widetilde{g}}=\xi^{D}\sqrt{-g}$. Hence, to obtain the correct coefficient of  $\hat R $,  meaning that $\xi^{5}\varphi \xi^{-2}=1$, we must choose $\xi=\varphi^{-1/3}$. Therefore, the conformal transformation of the metric (\ref{kkmetric}) becomes
\begin{eqnarray}\label{kkmetric2}
\widetilde{\hat g}_{MN}=\left(\begin{array}{r|r}
\varphi^{-2/3}g_{\mu\nu}+\varphi^{4/3}A_{\mu}A_{\nu}&\varphi^{4/3} A_{\mu}
\\[0.3em] \hline  \rule{0pt}{1.0\normalbaselineskip}
\varphi^{4/3} A_{\nu}&\varphi^{4/3}
\end{array}\right).
\end{eqnarray}
Inspired by equation (\ref{kkmetric2}), one can write a generalized KK metric ansatz, which is the special case of the DeWitt ansatz\cite{Cvetic:2003jy}, in terms of the actual 4D fields as follows:
\begin{eqnarray}\label{kkmetric3}
{\hat g}_{MN}=\left(\begin{array}{r|r}
e^{2\alpha\psi}g_{\mu\nu}+e^{2\beta\psi}A_{\mu}A_{\nu}&e^{2\beta\psi} A_{\mu}
\\[0.3em] \hline  \rule{0pt}{1.0\normalbaselineskip}
e^{2\beta\psi} A_{\nu}&e^{2\beta\psi}
\end{array}\right),
\end{eqnarray}
where the $\psi=\psi(x)$ is a new scalar field, and  $\alpha$ and $\beta$ are arbitrary constants, which will be determined later. The obvious choice of the vielbein basis is inspired by coframes  (\ref{coframes1}) as (see, e.g.,\cite{Pope,Perry})
\begin{eqnarray}
\hat E^{\hat \mu}(x,y)= e^{\alpha\psi(x)}E^{\hat \mu}(x),\qquad\qquad\qquad\qquad\hat E^{\hat 5}(x,y)= e^{\beta\psi(x)}\left[A(x)+dy\right].
\end{eqnarray}
By using the relation between dual fields,~$\hat E^{\hat B}(\hat X_{\hat A})=\hat\iota_{\hat X_{\hat A}}\hat E^{\hat B}=\hat\delta_{\hat A}\,^{\hat B}$, the basis vectors are explicitly found to be
\begin{eqnarray}\label{inners2}
\hat\iota_{\hat X_{\hat\mu}}(x,y)=e^{-\alpha\psi(x)}\left[\iota_{ X_{\hat\mu}}-{A}_{\hat\mu}(x)\iota_{ X_{y}}\right],\qquad\qquad\qquad
\hat\iota_{\hat X_{\hat5}}(x,y)=e^{-\beta\psi(x)}\iota_{ X_{y}},
\end{eqnarray}
where the ~$\iota_{ X_{\hat\mu}}E^{\hat\nu}=\delta_{\hat\mu}\,^{\hat\nu}$ is also satisfied in four dimensions. We can employ the horizontal lift basis formalism, which is the easiest way to obtain  not only field equations but also invariants. For this purpose, the 5D connection coefficients can be written as follows\cite{Misner:1973prb}:
 \begin{eqnarray}\label{connection2}
\hat\Gamma_{\hat A\hat B\hat C}=\frac{1}{2}\left[ \hat\iota_{\hat X_{\hat C}}\hat g_{\hat A\hat B}+
\hat\iota_{\hat X_{\hat B}}\hat g_{\hat A\hat C} - \hat\iota_{\hat X_{\hat A}}\hat
g_{\hat B\hat C}+\hat f_{\hat A\hat B\hat C}
+\hat f_{\hat A\hat C\hat B}+\hat f_{\hat C\hat B\hat A}\right].
\end{eqnarray}
Here, the commutation coefficients, $\hat f_{\hat A\hat B}\,^{\hat C}=-\hat f_{\hat B\hat A}\,^{\hat C}$, are evaluated by $[\hat\iota_{\hat X_{\hat A}},\hat\iota_{\hat X_{\hat B}}]\equiv\hat f_{\hat A\hat B}\,^{\hat C} \hat\iota_{\hat X_{\hat C}}$, and the block diagonal metric becomes $\hat g_{\hat A\hat B}=diag(g_{\hat\mu\hat\nu},+1)$. The nonzero commutators of the anholonomic basis vectors (\ref{inners2})
now take the form
\begin{eqnarray}\label{commu2}
&&\hat f_{\hat\mu\hat\nu}\,^{\hat\lambda}=-\alpha e^{-\alpha\psi}(\psi_{\hat\mu}\delta_{\hat\nu}\,^{\hat\lambda}
-\psi_{\hat\nu}\delta_{\hat\mu}\,^{\hat\lambda}),
\nonumber\\
&&
\hat f_{\hat\mu\hat\nu}\,^{\hat5}=-e^{(\beta-2\alpha)\psi} F_{\hat\mu\hat\nu},
\nonumber\\
&&
\hat f_{\hat\mu\hat5}\,^{\hat5}
=-\hat f_{\hat5\hat\mu}\,^{\hat5}=-\beta e^{-\alpha\psi}\psi_{\hat\mu}.
\end{eqnarray}
Hence, the required  higher-dimensional components of the connection are found by considering (\ref{connection2}) and (\ref{commu2}):
\begin{eqnarray}\label{anholomoniccristofel1}
&&\hat\Gamma^{\hat\mu}\,_{\hat\nu\hat\lambda}= e^{-\alpha\psi}\Gamma^{\hat\mu}\,_{\hat\nu\hat\lambda}
-\alpha e^{-\alpha\psi}(\psi^{\hat\mu}g_{\hat\nu\hat\lambda}
-\psi_{\hat\nu}\delta^{\hat\mu}\,_{\hat\lambda}),
\nonumber\\
&&
\hat\Gamma^{\hat\mu}\,_{\hat5\hat\nu}
=\hat\Gamma^{\hat\mu}\,_{\hat\nu\hat5}=-\frac{1}{2}e^{(\beta-2\alpha)\psi}   F^{\hat\mu}\,_{\hat\nu},
\nonumber\\
&&\hat\Gamma^{\hat5}\,_{\hat\mu\hat\nu}= \frac{1}{2} e^{(\beta-2\alpha)\psi}  F_{\hat\mu\hat\nu},
\qquad\quad
\hat\Gamma^{\hat\mu}\,_{\hat5\hat5}=-\beta e^{-\alpha\psi}\psi^{\hat\mu},
\qquad\quad
\hat\Gamma^{\hat5}\,_{\hat\mu\hat5}=\beta e^{-\alpha\psi}\psi_{\hat\mu}.
\end{eqnarray}
Conversely, in the noncoordinate basis the Riemann tensor is defined by Misner et al.\cite{Misner:1973prb}
\begin{eqnarray}\label{rie1}
\hat R^{\hat A}\,_{\hat B\hat C\hat D}=\hat\iota_{\hat X_{\hat C}}\hat\Gamma^{\hat A}\,_{\hat B\hat D} -\hat\iota_{\hat X_{\hat D}}\hat\Gamma^{\hat A}\,_{\hat B\hat C}+\hat\Gamma^{\hat A}\,_{\hat E
\hat C}\hat\Gamma^{\hat E}\,_{\hat B\hat D} -\hat\Gamma^{\hat A}\,_{\hat E\hat D}\hat\Gamma^{\hat E}\,_{\hat B\hat C}-\hat\Gamma^{\hat A}\,_{\hat B\hat E}\hat f_{\hat C\hat D}\,^{\hat E}.
\end{eqnarray}
Now, the dimensionally reduced vielbein components of the curvature tensor can be obtained using (\ref{inners2}), (\ref{commu2}), and (\ref{anholomoniccristofel1}) in (\ref{rie1})
\begin{eqnarray}\label{riem9}
&&\hat{R}^{\hat\mu}\,_{\hat\nu\hat\lambda\hat\sigma}=e^{-2\alpha\psi}R^{\hat\mu}\,_{\hat\nu\hat\lambda\hat\sigma}
-\frac{1}{4}e^{2(\beta-2\alpha)\psi}\left(2 F^{\hat\mu}\,_{\hat\nu}F_{\hat\lambda\hat\sigma}+F^{\hat\mu}\,_{\hat\lambda}F_{\hat\nu\hat\sigma}
-F^{\hat\mu}\,_{\hat\sigma}F_{\hat\nu\hat\lambda}\right)
\nonumber\\&&\qquad\qquad
+\alpha e^{-2\alpha\psi}\bigl(g_{\hat\nu\hat\lambda}D_{\hat\sigma}\psi^{\hat\mu}
-g_{\hat\nu\hat\sigma}D_{\hat\lambda}\psi^{\hat\mu}
+\delta^{\hat\mu}\,_{\hat\sigma}D_{\hat\lambda}\psi_{\hat\nu}
-\delta^{\hat\mu}\,_{\hat\lambda}D_{\hat\sigma}\psi_{\hat\nu}\bigr)
\nonumber\\&&\qquad\qquad
+\alpha^{2} e^{-2\alpha\psi}\Bigl[\psi^{\hat\mu}\bigl(\psi_{\hat\lambda}g_{\hat\nu\hat\sigma}
-\psi_{\hat\sigma}g_{\hat\nu\hat\lambda}\bigr)
+\psi_{\hat\rho}\psi^{\hat\rho}\bigl(g_{\hat\nu\hat\lambda}\delta^{\hat\mu}\,_{\hat\sigma}
-g_{\hat\nu\hat\sigma}\delta^{\hat\mu}\,_{\hat\lambda}\bigr)\nonumber\\
&&\qquad\qquad
+\psi_{\hat\nu}\bigl(\psi_{\hat\sigma}\delta^{\hat\mu}\,_{\hat\lambda}
-\psi_{\hat\lambda}\delta^{\hat\mu}\,_{\hat\sigma}\bigr)\Bigr],
\\
&&\hat{R}^{\hat\mu}\,_{\hat5\hat\lambda\hat\sigma}=-\frac{1}{2}e^{(\beta-3\alpha)\psi} D^{\hat\mu}F_{\hat\lambda\hat\sigma}
+\frac{1}{2}(\alpha-\beta)e^{(\beta-3\alpha)\psi} \left( 2\varphi^{\hat\mu} F_{\hat\lambda\hat\sigma}
+\varphi_{\hat\lambda} F^{\hat\mu}\,_{\hat\sigma}-\varphi_{\hat\sigma}F^{\hat\mu}\,_{\hat\lambda}\right)
\nonumber\\&&\qquad\qquad
+\frac{1}{2}\alpha e^{(\beta-3\alpha)\psi} \psi_{\hat\rho}\bigl(F^{\hat\rho}\,_{\hat\lambda}\delta^{\hat\mu}\,_{\hat\sigma}
-F^{\hat\rho}\,_{\hat\sigma}\delta^{\hat\mu}\,_{\hat\lambda}\bigr)
,\label{riem12}\\
&&\hat{R}^{\hat\mu}\,_{\hat5\hat\lambda\hat5}=-\beta e^{-2\alpha\psi}D_{\hat\lambda}\varphi^{\hat\mu}
-\frac{1}{4}e^{2(\beta-2\alpha)\psi} F^{\hat\mu\hat\rho}F_{\hat\rho\hat\lambda}
\nonumber\\&&\qquad\qquad
+e^{-2\alpha\psi}\left[\left(2\alpha-\beta\right)\beta\psi^{\hat\mu}\psi_{\hat\lambda}
-\alpha\beta\psi^{\hat\rho}\psi_{\hat\rho}\delta^{\hat\mu}\,_{\hat\lambda}
\right].\label{riem13}
\end{eqnarray}
After introducing ${\hat R}_{\hat\mu\hat\nu}\equiv\mathcal{\widetilde{P}}_{\hat\mu\hat\nu}$ and
${{\hat R}}_{\hat5\hat\nu}\equiv\mathcal{\widetilde{Q}}_{\hat\nu}$ together with
$\hat R_{\hat5\hat5}\equiv\mathcal{\widetilde{U}}$ and finding the vielbeins and
the inverse vielbeins as
\begin{eqnarray}\label{viel1}
\hat h^{\hat A}\,_{M} =\left(
\begin{array}{cc}
e^{\alpha\psi} h^{\hat\mu}\,_{\nu}   & 0 \\
e^{\beta\psi} A_{\nu} & e^{\beta\psi}
\end{array}
\right),
\qquad\qquad
\hat h^{M}\,_{\hat A}=\left(
\begin{array}{cc}
 e^{-\alpha\psi} h^\nu\,_{\hat\mu}   & 0 \\
- e^{-\alpha\psi}A_{\hat\mu} & e^{-\beta\psi}
\end{array}
\right),
\end{eqnarray}
we obtain the following components of the Ricci tensor in the coordinate basis
\begin{eqnarray}\label{}
&&{\hat R}_{\mu\nu}=e^{2\alpha\psi}\mathcal{\widetilde{P}}_{\mu\nu}
+e^{(\alpha+\beta)\psi} A_{\mu}\mathcal{\widetilde{Q}}_{\nu}
+e^{(\alpha+\beta)\psi} A_{\nu}\mathcal{\widetilde{Q}}_{\mu}
+e^{2\beta\psi}A_{\mu}A_{\nu}\mathcal{\widetilde{U}}
 ,\nonumber\\
&&{{\hat R}}_{5\nu}=e^{(\alpha+\beta)\psi}\mathcal{\widetilde{Q}}_{\nu}
+e^{2\beta\psi}A_{\nu}\mathcal{\widetilde{U}}
 ,\nonumber\\
&&\hat R_{55}=e^{2\beta\psi}\mathcal{\widetilde{U}},
\end{eqnarray}
where
\begin{eqnarray}\label{newricci}
&&\mathcal{\widetilde{P}}_{\mu\nu}
=e^{-2\alpha\psi}R_{\mu\nu}
-\frac{1}{2}e^{2(\beta-2\alpha)\psi}
F_{\mu\rho}F_{\nu}\,^{\rho}
- e^{-2\alpha\psi}\left[ \alpha g_{\mu\nu}D_{\rho}\psi^{\rho}
+\left(2\alpha+\beta\right)D_{\mu}\psi_{\nu}\right]
\nonumber\\&&\qquad\quad
-e^{-2\alpha\psi}\left[(2\alpha+\beta)\alpha g_{\mu\nu}\psi_{\rho}\psi^{\rho}
-\left(2\alpha^{2}+2\alpha\beta-\beta^{2}\right)\psi_{\mu}\psi_{\nu}\right]
,\nonumber\\
&&\mathcal{\widetilde{Q}}_{\nu}
=-\frac{1}{2} e^{(\beta-3\alpha)\psi}{D}_{\rho}F^{\rho}\,_{\nu}
-\frac{3}{2}\beta e^{(\beta-3\alpha)\psi}\psi_{\rho}F^{\rho}\,_{\nu} ,\nonumber\\
&&\mathcal{\widetilde{U}}
=-\beta e^{-2\alpha\psi}D_{\rho}\psi^{\rho}+\frac{1}{4}e^{2(\beta-2\alpha)\psi} F_{\rho\tau}F^{\rho\tau}-(2\alpha+\beta)\beta e^{-2\alpha\psi}\psi_{\rho}\psi^{\rho}.
\end{eqnarray}
Finally, one easily finds the solution set of the conformal KK theory as
$\widetilde{S}_{KK}=\{\mathcal{\widetilde{P}}_{\mu\nu}=0,~\mathcal{\widetilde{Q}}_{\nu}=0,
~\mathcal{\widetilde{U}}=0\}$. Next, we can introduce that
\begin{eqnarray}
\hat{D}_{\hat A}\hat{R}^{\hat A}\,_{\hat\nu\hat\lambda\hat\sigma}\equiv\widetilde{\mathbb{P}}_{\hat\nu\hat\lambda\hat\sigma},\quad
\hat{D}_{\hat A}\hat{R}^{\hat A}\,_{\hat5\hat\lambda\hat\sigma}\equiv\widetilde{\mathbb{S}}_{\hat\lambda\hat\sigma},\quad
\hat{D}_{\hat A}\hat{R}^{\hat A}\,_{\hat\nu\hat\lambda\hat5}\equiv\widetilde{\mathbb{Q}}_{\hat\nu\hat\lambda},\quad
\hat{D}_{\hat A}\hat{R}^{\hat A}\,_{\hat5\hat\lambda \hat5}\equiv\widetilde{\mathbb{U}}_{\hat\lambda}.
\end{eqnarray}
Then, we can derive the desired field equations in coordinate basis, as mentioned before, by substituting the connections (\ref{anholomoniccristofel1}) together with the $\{\mathcal{\widetilde{P}}_{\hat\mu\hat\nu},\mathcal{\widetilde{Q}}_{\hat\nu},\mathcal{\widetilde{U}}\}$~set
in (\ref{newricci}), and use the vielbeins fields (\ref{viel1}) in the following forms:
\begin{eqnarray}
&&\hat{D}_{A}\hat{R}^{A}\,_{\nu\lambda\sigma}=e^{3\alpha\psi}\widetilde{\mathbb{P}}_{\nu\lambda\sigma}
+e^{(2\alpha+\beta)\psi} A_{\nu}\widetilde{\mathbb{S}}_{\lambda\sigma}
+e^{(2\alpha+\beta)\psi} A_{\sigma}\widetilde{\mathbb{Q}}_{\nu\lambda}
-e^{(2\alpha+\beta)\psi} A_{\lambda}\widetilde{\mathbb{Q}}_{\nu\sigma}
\nonumber\\
&&\qquad\qquad\qquad
+e^{(\alpha+2\beta)\psi}A_{\nu}A_{\sigma}\widetilde{\mathbb{U}}_{\lambda}
-e^{(\alpha+2\beta)\psi}A_{\nu}A_{\lambda}\widetilde{\mathbb{U}}_{\sigma},\nonumber\\
&&\hat{D}_{A}\hat{R}^{A}\,_{5\lambda\sigma}=e^{(2\alpha+\beta)\psi}\widetilde{\mathbb{S}}_{\lambda\sigma}
+e^{(\alpha+2\beta)\psi}A_{\sigma}\widetilde{\mathbb{U}}_{\lambda}
-e^{(\alpha+2\beta)\psi}A_{\lambda}\widetilde{\mathbb{U}}_{\sigma},\nonumber\\
&&\hat{D}_{A}\hat{R}^{A}\,_{\nu\lambda5}=-\hat{D}_{A}\hat{R}^{A}\,_{\nu5\lambda}
=e^{(2\alpha+\beta)\psi}\widetilde{\mathbb{Q}}_{\nu\lambda}
+e^{(\alpha+2\beta)\psi}A_{\nu}\widetilde{\mathbb{U}}_{\lambda},\nonumber\\
&&\hat{D}_{A}\hat{R}^{A}\,_{5\lambda 5}=-\hat{D}_{A}\hat{R}^{A}\,_{55\lambda}=e^{(\alpha+2\beta)\psi}\widetilde{\mathbb{U}}_{\lambda},
\end{eqnarray}
where
\begin{eqnarray}
 & &\widetilde{\mathbb{P}}_{\nu\lambda\sigma}= e^{-\alpha\psi}\bigg\{D_{\lambda}\mathcal{\widetilde{P}}_{\nu \sigma}-D_{\sigma}\mathcal{\widetilde{P}}_{\nu \lambda}
+ \frac{1}{2}e^{(\beta-\alpha)\psi} \left( F_{\nu \sigma}\mathcal{\widetilde{Q}}_{\lambda}-F_{\nu \lambda}\mathcal{\widetilde{Q}}_{\sigma}
+ 2 F_{\lambda\sigma}\mathcal{\widetilde{Q}}_{\nu} \right)\nonumber\\
&&\qquad\quad
+\alpha\left[\psi^{\rho}\left(g_{\nu\lambda}\mathcal{\widetilde{P}}_{\rho \sigma}
-g_{\nu\sigma}\mathcal{\widetilde{P}}_{\rho \lambda}\right)
+\psi_{\lambda}\mathcal{\widetilde{P}}_{\nu \sigma}-\psi_{\sigma}\mathcal{\widetilde{P}}_{\nu \lambda}\right]\bigg\}
,\label{p11}  \\
& & \widetilde{\mathbb{S}}_{\lambda\sigma}=e^{-\alpha\psi}\bigg[D_{\lambda}\mathcal{\widetilde{Q}}_{\sigma}
-D_{\sigma}\mathcal{\widetilde{Q}}_{\lambda}- \frac{1}{2}e^{(\beta-\alpha)\psi} \left(F_{\lambda}\,^{\rho}\mathcal{\widetilde{P}}_{\rho\sigma}-F_{\sigma}\,^{\rho}
\mathcal{\widetilde{P}}_{\rho\lambda}\right)
+e^{(\beta-\alpha)\psi} F_{\lambda\sigma}\mathcal{\widetilde{U}}\nonumber\\
&&\qquad\quad
+\alpha\left(\psi_{\lambda}\mathcal{\widetilde{Q}}_{\sigma}
-\psi_{\sigma}\mathcal{\widetilde{Q}}_{\lambda}\right)
\bigg],\label{p22}   \\
 & & \widetilde{\mathbb{Q}}_{\nu\lambda}=e^{-\alpha\psi}\bigg[D_{\lambda}\mathcal{\widetilde{Q}}_{\nu}
 +\alpha g_{\nu\lambda}\psi^{\rho} \mathcal{\widetilde{Q}}_{\rho}
 + (\beta-\alpha)\psi_{\nu} \mathcal{\widetilde{Q}}_{\lambda}
 +  \beta\psi_{\lambda}\mathcal{\widetilde{Q}}_{\nu}
 +\frac{1}{2}e^{(\beta-\alpha)\psi}   F_{\nu}\,^{\rho} \mathcal{\widetilde{P}}_{\rho\lambda}\nonumber\\
&&\qquad\quad
 -\frac{1}{2}e^{(\beta-\alpha)\psi} F_{\nu\lambda}\mathcal{\widetilde{U}}\bigg], \label{p33}  \\
 & & \widetilde{\mathbb{U}}_{\lambda}=e^{-\alpha\psi}\left(D_{\lambda}\mathcal{\widetilde{U}}
 +\beta\psi_{\lambda}\mathcal{\widetilde{U}}
-\frac{1}{2} e^{(\beta-\alpha)\psi} F_{\lambda}\,^{\rho}\mathcal{\widetilde{Q}}_{\rho}
- \beta\psi^{\rho} \mathcal{\widetilde{P}}_{\rho\lambda}\right).\label{p44}
\end{eqnarray}
In this  regard, we have the solution set: $\widetilde{S}_{WYKK}=\{\widetilde{\mathbb{P}}_{\nu\lambda\sigma}=0,\;~\widetilde{\mathbb{S}}_{\lambda\sigma}=0,
\;~\widetilde{\mathbb{Q}}_{\nu\lambda}=0,\; \widetilde{\mathbb{U}}_{\lambda}=0 \}$. Again, it is easy to see that $\widetilde{\mathbb{S}}_{\lambda\sigma}=\widetilde{\mathbb{Q}}_{\sigma\lambda}-\widetilde{\mathbb{Q}}_{\lambda\sigma}$, so that we can ignore  equation (\ref{p22}) without loss of generality. Hence, the final forms of the reduced field equations of the transformed WYKK theory become
\begin{eqnarray}\label{generalf}
 & &D_{\lambda}\mathcal{\widetilde{P}}_{\nu \sigma}-D_{\sigma}\mathcal{\widetilde{P}}_{\nu \lambda}
+ \frac{1}{2}e^{(\beta-\alpha)\psi} \left( F_{\nu \sigma}\mathcal{\widetilde{Q}}_{\lambda}-F_{\nu \lambda}\mathcal{\widetilde{Q}}_{\sigma}
+ 2 F_{\lambda\sigma}\mathcal{\widetilde{Q}}_{\nu} \right)
+\alpha\Big[\psi^{\rho}\left(g_{\nu\lambda}\mathcal{\widetilde{P}}_{\rho \sigma}
-g_{\nu\sigma}\mathcal{\widetilde{P}}_{\rho \lambda}\right)
\nonumber\\
&&
+\psi_{\lambda}\mathcal{\widetilde{P}}_{\nu \sigma}-\psi_{\sigma}\mathcal{\widetilde{P}}_{\nu \lambda}\Big]=0
,\label{p14}\nonumber  \\
 & & D_{\lambda}\mathcal{\widetilde{Q}}_{\nu}
 +\alpha g_{\nu\lambda}\psi^{\rho} \mathcal{\widetilde{Q}}_{\rho}
 + (\beta-\alpha)\psi_{\nu} \mathcal{\widetilde{Q}}_{\lambda}
 +  \beta\psi_{\lambda}\mathcal{\widetilde{Q}}_{\nu}
 +\frac{1}{2}e^{(\beta-\alpha)\psi}   F_{\nu}\,^{\rho} \mathcal{\widetilde{P}}_{\rho\lambda}
 -\frac{1}{2}e^{(\beta-\alpha)\psi} F_{\nu\lambda}\mathcal{\widetilde{U}}=0, \label{p34} \nonumber \\
 & & D_{\lambda}\mathcal{\widetilde{U}}
 +\beta\psi_{\lambda}\mathcal{\widetilde{U}}
-\frac{1}{2} e^{(\beta-\alpha)\psi} F_{\lambda}\,^{\rho}\mathcal{\widetilde{Q}}_{\rho}
- \beta\psi^{\rho} \mathcal{\widetilde{P}}_{\rho\lambda}=0.\label{p44}
\end{eqnarray}
As expected,  the corresponding equations of motion are
more complicated than those given in (\ref{proper1})$-$(\ref{proper3}).

\section{The Reduced Actions from the Transformed WYKK Theory}
\label{sec4}
By considering a popular dimensional reduction method, the quadratic curvature term can be expanded as
\begin{eqnarray}\label{expr2}
{\mathcal{\hat {K}}}={\hat {R}}_{ABCD}{\hat {R}}^{ABCD}=\hat{R}_{\hat\mu\hat\nu\hat\lambda\hat\sigma}\hat{R}^{\hat\mu\hat\nu\hat\lambda\hat\sigma}
+4\hat{R}_{\hat\mu\hat5\hat\lambda\hat\sigma}\hat{R}^{\hat\mu\hat5\hat\lambda\hat\sigma}
+4\hat{R}_{\hat\mu\hat5\hat\lambda\hat5}\hat{R}^{\hat\mu\hat5\hat\lambda\hat5}.
\end{eqnarray}
As is well-known, the invariant does not depend on the choice of basis. Thus, by using directly (\ref{riem9})$-$(\ref{riem13}) and the relation $2R_{\mu\nu\lambda\sigma}F^{\mu\lambda}F^{\nu\sigma}=R_{\mu\nu\lambda\sigma}F^{\mu\nu}F^{\lambda\sigma}$ and
after long but careful manipulations and reorganization of the terms, we have the following reduced expression:
\begin{eqnarray}\label{inv2}
&&
\mathcal{\hat {K}}=e^{-4\alpha\psi}\Big[R_{\mu\nu\lambda\sigma}R^{\mu\nu\lambda\sigma}-\frac{3}{2}e^{2(\beta-\alpha)\psi} R_{\mu\nu\lambda\sigma}F^{\mu\nu}F^{\lambda\sigma}
+\frac{3}{8}e^{4(\beta-\alpha)\psi}F_{\mu\nu}F^{\mu\nu}F_{\lambda\sigma}F^{\lambda\sigma}
\nonumber\\&&\quad
+\frac{5}{8}e^{4(\beta-\alpha)\psi}F_{\mu\nu}F^{\nu\lambda}F_{\lambda\sigma}F^{\sigma\mu}
+(9\alpha^{2}-14\alpha\beta+6\beta^{2})e^{2(\beta-\alpha)\psi}\psi_{\lambda}\psi^{\lambda} F_{\mu\nu}F^{\mu\nu}
\nonumber\\&&\quad
+2(3\alpha^{2}-2\alpha\beta-2\beta^{2})e^{2(\beta-\alpha)\psi}\psi_{\mu}\psi^{\lambda}F_{\lambda\nu}F^{\nu\mu}
+e^{2(\beta-\alpha)\psi}D_{\lambda}F_{\mu\nu}D^{\lambda}F^{\mu\nu}
\nonumber\\&&\quad
+ 4(\beta-\alpha)e^{2(\beta-\alpha)\psi} \left( \psi^{\lambda} F^{\mu\nu}+\psi^{\mu} F^{\lambda\nu}\right)D_{\lambda}F_{\mu\nu}
+ 2(3\alpha-\beta)e^{2(\beta-\alpha)\psi} F^{\mu\lambda}F^{\nu}\,_{\lambda}D_{\mu}\psi_{\nu}
\nonumber\\&&\quad
+4(2\alpha^{2}+\beta^{2})D_{\mu}\psi_{\nu}D^{\mu}\psi^{\nu}
+4\alpha e^{2(\beta-\alpha)\psi}\psi^{\mu} F_{\mu\nu}D_{\lambda}F^{\lambda\nu}
+8\alpha(\alpha\psi_{\mu}\psi_{\nu}-D_{\mu}\psi_{\nu})R^{\mu\nu}
\nonumber\\&&\quad
-4\alpha^{2}\psi_{\mu}\psi^{\mu}R
+4\alpha^{2}D_{\mu}\psi^{\mu}D_{\nu}\psi^{\nu}
+8\alpha(2\alpha^{2}+\beta^{2})\psi_{\mu}\psi^{\mu} D_{\nu}\psi^{\nu}
\nonumber\\&&\quad
-8(2\alpha^{3}+2\alpha\beta^{2}-\beta^{3})\psi^{\mu}\psi^{\nu} D_{\mu}\psi_{\nu}
+4(3\alpha^{4}+4\alpha^{2}\beta^{2}-2\alpha\beta^{3}+\beta^{4})\psi_{\mu}\psi^{\mu}\psi_{\nu}\psi^{\nu}
\Big],
\end{eqnarray}
where the 4D Ricci tensor and Ricci scalar appear in the above conformal equation. Furthermore, the comparison between the invariants (\ref{inv1}) and (\ref{inv2}) causes  us to introduce new interaction terms such as
$\psi^{2} R$,  $\psi^{2} D\psi$, and $\psi^{4}$.

Another way of obtaining the reduced Kretschmann scalar (\ref{inv2}) is directly using the conformal transformation rule of the squaring curvature $\widetilde{R}_{\mu\nu\lambda\sigma}\widetilde{R}^{\mu\nu\lambda\sigma}$, which is expressed as (see, e.g.,\cite{Carneiro:2004rt,Bao:2007fx})
\begin{eqnarray}\label{coninv}
&&\widetilde{R}_{\mu\nu\lambda\sigma}\widetilde{R}^{\mu\nu\lambda\sigma}
=e^{-4\xi}\Big\{R_{\mu\nu\lambda\sigma}R^{\mu\nu\lambda\sigma}
+8\left(\xi_{\mu}\xi_{\nu}-D_{\mu}\xi_{\nu}\right)R^{\mu\nu}
+4\big[D_{\mu}\xi^{\mu}D_{\nu}\xi^{\nu}
\nonumber\\&&\qquad\qquad\qquad\quad
+(D-2)D_{\mu}\xi_{\nu}D^{\mu}\xi^{\nu}\big]
-4\xi_{\mu}\xi^{\mu}R
+8(D-2)\big[\xi_{\mu}\xi^{\mu}D_{\nu}\xi^{\nu}
-\xi^{\mu}\xi^{\nu}D_{\mu}\xi_{\nu}\big]
\nonumber\\&&\qquad\qquad\qquad\quad
+2(D-1)(D-2)\xi_{\mu}\xi^{\mu}\xi_{\nu}\xi^{\nu}\Big\},
\end{eqnarray}
if the metric changes as
\begin{eqnarray}\label{}
\widetilde{g}_{\mu\nu}(x)=e^{2\xi(x)}g_{\mu\nu}(x).
\end{eqnarray}
Conversely, the 5D metric (\ref{kkmetric3}) can be rewritten in terms of 4D fields as follows:
\begin{eqnarray}\label{kkmetric5}
\widetilde{\hat {g}}_{MN}=e^{2\alpha\psi}\left(\begin{array}{r|r}
g_{\mu\nu}+e^{2(\beta-\alpha)\psi}A_{\mu}A_{\nu}&e^{2(\beta-\alpha)\psi} A_{\mu}
\\[0.3em] \hline  \rule{0pt}{1.0\normalbaselineskip}
e^{2(\beta-\alpha)\psi} A_{\nu}&e^{2(\beta-\alpha)\psi}
\end{array}\right),
\end{eqnarray}
which means that the new conformal factor is equal to $\xi(x)=\alpha\psi(x)$. Hence, equation (\ref{coninv}) changes the following result for  $D=5$
\begin{eqnarray}\label{confinv}
&&\widetilde{\hat {R}}_{ABCD}\widetilde{\hat {R}}^{ABCD}
=e^{-4\alpha\psi}\bigg\{\hat R_{ABCD}\hat R^{ABCD}
+8\alpha\left(\alpha\hat\psi_{A}\hat\psi_{B}-\hat D_{A}\hat\psi_{B}\right)\hat R^{AB}
\nonumber\\&&\qquad\qquad\qquad\quad~
-4\alpha^{2}\hat\psi_{A}\hat\psi^{A}\hat R
+4\alpha^{2}\left[\hat D_{A}\hat\psi^{A}\hat D_{B}\hat\psi^{B}+3\hat D_{A}\hat\psi_{B}\hat D^{A}\hat\psi^{B}\right]
\nonumber\\&&\qquad\qquad\qquad\quad~
+24\alpha^{3}\left[\hat\psi_{A}\hat\psi^{A}\hat D_{B}\hat\psi^{B}
-\hat\psi^{A}\hat\psi^{B}\hat D_{A}\hat\psi_{B}\right]
+24\alpha^{4}\hat\psi_{A}\hat\psi^{A}\hat\psi_{B}\hat\psi^{B}\bigg\}.
\end{eqnarray}
Moreover, a comparison between  metrics (\ref{kkmetric}) and (\ref{kkmetric5}) leads us to obtain the new potential term, $e^{(\beta-\alpha)\psi(x)}$. As a result,  terms $\hat R_{ABCD}\hat R^{ABCD}$ in (\ref{inv1}), the anholonomic components of  $\hat R^{\hat A\hat B}$, and $\hat R$, which comes from  equation (\ref{riem}), rescale under the redefined scalar field, $\varphi(x)= e^{(\beta-\alpha)\psi(x)}$, transformation. For example, the curvature invariant changes accordingly in the following form:
\begin{eqnarray}\label{changecurv}
\hat R=R
-2(\beta-\alpha)^{2}\psi_{\mu}\psi^{\mu}
-\frac{1}{4}e^{2(\beta-\alpha)\psi}F_{\mu\nu}F^{\mu\nu}
-2(\beta-\alpha)D_{\mu}\psi^{\mu}.
\end{eqnarray}
We should also calculate the following necessary relations of the function $\psi(x)$ as
\begin{eqnarray}\label{relations}
&&\hat{D}_{\hat\mu}\psi_{\hat\nu}={D}_{\hat\mu}\psi_{\hat\nu},\nonumber\\&&
\hat{D}_{\hat\mu}\psi_{\hat5}=\hat{D}_{\hat5}\psi_{\hat\mu}=
\frac{1}{2}e^{(\beta-\alpha)\psi}\psi_{\hat\lambda}F^{\hat\lambda}\,_{\hat\mu},
\nonumber\\&&
\hat{D}_{\hat5}\psi_{\hat5}=(\beta-\alpha)\psi_{\hat\lambda}\psi^{\hat\lambda},
\end{eqnarray}
and it is clear that $\psi_{\hat5}=\psi^{\hat5}=0$, by considering  the new versions of the connections (\ref{anholomoniccristofel}), which are obtained  the same way
\begin{eqnarray}\label{anholomoniccristofel2}
&&\hat\Gamma^{\hat\mu}\,_{\hat\nu\hat\lambda}=\Gamma^{\hat\mu}\,_{\hat\nu\hat\lambda},
\qquad\qquad\qquad
\hat\Gamma^{\hat5}\,_{\hat\mu\hat\nu}= \frac{1}{2} e^{(\beta-\alpha)\psi}F_{\hat\mu\hat\nu},
\nonumber\\&&
\hat\Gamma^{\hat\mu}\,_{\hat5\hat\nu}
=\hat\Gamma^{\hat\mu}\,_{\hat\nu\hat5}=-\frac{1}{2}e^{(\beta-\alpha)\psi} F^{\hat\mu}\,_{\hat\nu},
\nonumber\\&&
\hat\Gamma^{\hat\mu}\,_{\hat5\hat5}=-(\beta-\alpha)\psi^{\hat\mu},
\qquad\qquad\qquad
\hat\Gamma^{\hat5}\,_{\hat\mu\hat5}=(\beta-\alpha)\psi_{\hat\mu}.
\end{eqnarray}
As a mathematical challenge, if we substitute all results into  equation (\ref{confinv}), after careful calculations, we exactly obtain equation (\ref{inv2}). Conversely, in the lower dimension, the modified KK action (\ref{invari1}) is transformed into
\begin{eqnarray}\label{act3}
&&\widetilde{\hat I}=\frac{1}{2\hat\kappa^{2}}\int_{\hat{\mathcal{M}}}d^{4}x\,dy\,\sqrt{-{ g}}
e^{(4\alpha+\beta)\psi}\,e^{-2\alpha\psi}
\nonumber\\&&
\times \left[R-2(3\alpha^{2}+2\alpha\beta+\beta^{2})\psi_{\mu}\psi^{\mu}
-\frac{1}{4}e^{2(\beta-\alpha)\psi}F_{\mu\nu}F^{\mu\nu}-2(3\alpha+\beta) D_{\mu}\psi^{\mu} \right],
\end{eqnarray}
 where we use $\hat R$ that is obtained by (\ref{riem9})$-$(\ref{riem13}), and
$\sqrt{-\hat{g}}=e^{\xi D}\varphi\sqrt{-g}=e^{(4\alpha+\beta)\psi}\sqrt{-g}$
for $D=5$. We can find exactly the same result by considering the rescaled curvature scalar in Einstein frame, which is once again given by Carneiro et al.\cite{Carneiro:2004rt} and Bao et al.\cite{Bao:2007fx}
\begin{eqnarray}\label{beren}
\widetilde{R}=e^{-2\xi}\Big[R-(D-1)(D-2)\xi_{\mu}\xi^{\mu}-2(D-1)D_{\mu}\xi^{\mu}
\Big],
\end{eqnarray}
together with the equations (\ref{changecurv}) and (\ref{relations}). This is the easy way, if one wants to obtain field equations from implementing the least action principle to the action after computing the Ricci scalar by only considering (\ref{beren}).

\section{The special cases}
\label{sec5}
It is easy to read off from the action
(\ref{act3}) that to rid of the coefficient of the Einstein$-$\linebreak Hilbert term, i.e., to obtain minimally coupled gravity, we must choose  $\beta=-2\alpha$.\linebreak  Besides, the term $\psi_{\mu}\psi^{\mu}$ can be a canonical normalized term if $\alpha^{2}=1/12$, i.e., $(\alpha,\beta)=(\pm1/2\sqrt{3},\mp1/\sqrt{3})$. Note that,  $(\alpha,\beta)=(-1/3,2/3)$ in (\ref{kkmetric2}) also satisfies the condition $\beta=-2\alpha$, and see\cite{Gibbons:1985ac} for $(\alpha,\beta)=(-1/\sqrt{3},2/\sqrt{3})$. Conversely, for a $D=N+1$ dimensional spacetime, the proper values of the constants become $\alpha^{2}=1/\left[2(N-1)(N-2)\right]$ and $\beta=-(N-2)\alpha$ in\cite{Pope}, meaning that $(\alpha,\beta)=(\pm1/2\sqrt{3},\mp1/\sqrt{3})$ is obviously correct for $N=4$. The $\hat R$ term in (\ref{act3}) is also compatible with the result of \cite{Pope} for $N=4$, except for the coefficient of the term $D_{\mu}\psi^{\mu}$, which is a total divergence and does not contribute to the field equations, as usual. However, we can say that it seems that there is a typographical error  in\cite{Pope} because the same coefficient, i.e., $-2(3\alpha+\beta)$,  was also obtained in\cite{Perry}. Finally, if we consider the modified WYKK action from  (\ref{inv2}), then we have
\begin{eqnarray}\label{lastaction}
&&\widetilde{\hat I}=\frac{1}{2\hat\kappa^{2}}\int_{\hat{\mathcal{M}}}d^{4}x\,dy\,\sqrt{-{ g}}
e^{(4\alpha+\beta)\psi}
\,e^{-4\alpha\psi}\Big[R_{\mu\nu\lambda\sigma}R^{\mu\nu\lambda\sigma}-\frac{3}{2}e^{2(\beta-\alpha)\psi} R_{\mu\nu\lambda\sigma}F^{\mu\nu}F^{\lambda\sigma}
\nonumber\\&&\qquad
+\frac{3}{8}e^{4(\beta-\alpha)\psi}F_{\mu\nu}F^{\mu\nu}F_{\lambda\sigma}F^{\lambda\sigma}
+\frac{5}{8}e^{4(\beta-\alpha)\psi}F_{\mu\nu}F^{\nu\lambda}F_{\lambda\sigma}F^{\sigma\mu}
\nonumber\\&&\qquad
+(9\alpha^{2}-14\alpha\beta+6\beta^{2})e^{2(\beta-\alpha)\psi}\psi_{\lambda}\psi^{\lambda} F_{\mu\nu}F^{\mu\nu}
\nonumber\\&&\qquad
+2(3\alpha^{2}-2\alpha\beta-2\beta^{2})e^{2(\beta-\alpha)\psi}\psi_{\mu}\psi^{\lambda}F_{\lambda\nu}F^{\nu\mu}
\nonumber\\&&\qquad
+e^{2(\beta-\alpha)\psi}D_{\lambda}F_{\mu\nu}D^{\lambda}F^{\mu\nu}
+ 4(\beta-\alpha)e^{2(\beta-\alpha)\psi} \left( \psi^{\lambda} F^{\mu\nu}+\psi^{\mu} F^{\lambda\nu} \right)D_{\lambda}F_{\mu\nu}
\nonumber\\&&\qquad
+ 2(3\alpha-\beta)e^{2(\beta-\alpha)\psi} F^{\mu\lambda}F^{\nu}\,_{\lambda}D_{\mu}\psi_{\nu}
+4(2\alpha^{2}+\beta^{2})D_{\mu}\psi_{\nu}D^{\mu}\psi^{\nu}
\nonumber\\&&\qquad
+4\alpha e^{2(\beta-\alpha)\psi}\psi^{\mu} F_{\mu\nu}D_{\lambda}F^{\lambda\nu}
+8\alpha(\alpha\psi_{\mu}\psi_{\nu}-D_{\mu}\psi_{\nu})R^{\mu\nu}
-4\alpha^{2}\psi_{\mu}\psi^{\mu}R
\nonumber\\&&\qquad
+4\alpha^{2}D_{\mu}\psi^{\mu}D_{\nu}\psi^{\nu}
+8\alpha(2\alpha^{2}+\beta^{2})\psi_{\mu}\psi^{\mu} D_{\nu}\psi^{\nu}
-8(2\alpha^{3}+2\alpha\beta^{2}-\beta^{3})\psi^{\mu}\psi^{\nu} D_{\mu}\psi_{\nu}
\nonumber\\&&\qquad
+4(3\alpha^{4}+4\alpha^{2}\beta^{2}-2\alpha\beta^{3}+\beta^{4})\psi_{\mu}\psi^{\mu}\psi_{\nu}\psi^{\nu}
\Big].
\end{eqnarray}
Now, we have to set $\beta=0$ to obtain a standard form of $R_{\mu\nu\lambda\sigma}R^{\mu\nu\lambda\sigma}$. Hence, the two special cases are investigated to find the new reduced equations in the following subsections:

\subsection{The $\beta=-2\alpha$ case}

For $\beta=-2\alpha$ case, the field equations (\ref{generalf}) are simplified as follows:
\begin{eqnarray}\label{lastfieldson}
 & &D_{\lambda}\mathcal{\widetilde{P}}_{\nu \sigma}-D_{\sigma}\mathcal{\widetilde{P}}_{\nu \lambda}
+ \frac{1}{2}e^{-3\alpha\psi} \left( F_{\nu \sigma}\mathcal{\widetilde{Q}}_{\lambda}-F_{\nu \lambda}\mathcal{\widetilde{Q}}_{\sigma}
+ 2 F_{\lambda\sigma}\mathcal{\widetilde{Q}}_{\nu} \right)
+\alpha\Big[\psi^{\rho}\left(g_{\nu\lambda}\mathcal{\widetilde{P}}_{\rho \sigma}
-g_{\nu\sigma}\mathcal{\widetilde{P}}_{\rho \lambda}\right)
\nonumber\\&&
+\psi_{\lambda}\mathcal{\widetilde{P}}_{\nu \sigma}-\psi_{\sigma}\mathcal{\widetilde{P}}_{\nu \lambda}\Big]=0
,\label{pw1} \nonumber \\
 & & D_{\lambda}\mathcal{\widetilde{Q}}_{\nu}
 +\alpha g_{\nu\lambda}\psi^{\rho} \mathcal{\widetilde{Q}}_{\rho}
-3\alpha\psi_{\nu} \mathcal{\widetilde{Q}}_{\lambda}
 -2\alpha\psi_{\lambda}\mathcal{\widetilde{Q}}_{\nu}
 +\frac{1}{2}e^{-3\alpha\psi}   F_{\nu}\,^{\rho} \mathcal{\widetilde{P}}_{\rho\lambda}
 -\frac{1}{2}e^{-3\alpha\psi} F_{\nu\lambda}\mathcal{\widetilde{U}}=0, \label{pw3} \nonumber \\
 & & D_{\lambda}\mathcal{\widetilde{U}}
 -2\alpha\psi_{\lambda}\mathcal{\widetilde{U}}
-\frac{1}{2} e^{-3\alpha\psi} F_{\lambda}\,^{\rho}\mathcal{\widetilde{Q}}_{\rho}
+2\alpha\psi^{\rho} \mathcal{\widetilde{P}}_{\rho\lambda}=0.\label{pw4}
\end{eqnarray}
Here, the conformal KK equations $\{\mathcal{\widetilde{P}}_{\mu\nu},~\mathcal{\widetilde{Q}}_{\nu},
~\mathcal{\widetilde{U}}\}$ become
\begin{eqnarray}\label{son1}
&&\mathcal{\widetilde{P}}_{\mu\nu}
=e^{-2\alpha\psi}R_{\mu\nu}
-\frac{1}{2}e^{-8\alpha\psi}
F_{\mu\rho}F_{\nu}\,^{\rho}
-\alpha e^{-2\alpha\psi}  g_{\mu\nu}D_{\rho}\psi^{\rho}
-6\alpha^{2}e^{-2\alpha\psi}\psi_{\mu}\psi_{\nu}
,\nonumber\\
&&\mathcal{\widetilde{Q}}_{\nu}
=-\frac{1}{2} e^{-5\alpha\psi}{D}_{\rho}F^{\rho}\,_{\nu}
+3\alpha e^{-5\alpha\psi}\psi_{\rho}F^{\rho}\,_{\nu} =-\frac{1}{2}e^{\alpha\psi}{D}_{\rho}(e^{-6\alpha\psi}F^{\rho}\,_{\nu} ),\nonumber\\
&&\mathcal{\widetilde{U}}
=2\alpha e^{-2\alpha\psi}D_{\rho}\psi^{\rho}+\frac{1}{4}e^{-8\alpha\psi} F_{\rho\tau}F^{\rho\tau},
\end{eqnarray}
from the 4D point of view. There is full agreement between the above equations (\ref{son1}) and \cite{Pope} for $N=4$. Conversely, for action (\ref{lastaction}) one finds that
\begin{eqnarray}\label{}
&&
\widetilde{\hat I}=\frac{1}{2\hat\kappa^{2}}\int_{\hat{\mathcal{M}}}d^{4}x\,dy\,\sqrt{-{ g}}
e^{-2\alpha\psi}
\,\Big[R_{\mu\nu\lambda\sigma}R^{\mu\nu\lambda\sigma}-\frac{3}{2}e^{-6\alpha\psi} R_{\mu\nu\lambda\sigma}F^{\mu\nu}F^{\lambda\sigma}
\nonumber\\&&\quad\quad
+\frac{3}{8}e^{-12\alpha\psi}F_{\mu\nu}F^{\mu\nu}F_{\lambda\sigma}F^{\lambda\sigma}
+\frac{5}{8}e^{-12\alpha\psi}F_{\mu\nu}F^{\nu\lambda}F_{\lambda\sigma}F^{\sigma\mu}
+61\alpha^{2}e^{-6\alpha\psi} \psi_{\lambda}\psi^{\lambda} F_{\mu\nu}F^{\mu\nu}
\nonumber\\&&\quad\quad
-2\alpha^{2}e^{-6\alpha\psi} \psi_{\mu}\psi^{\lambda}F_{\lambda\nu}F^{\nu\mu}
+e^{-6\alpha\psi} D_{\lambda}F_{\mu\nu}D^{\lambda}F^{\mu\nu}
- 12\alpha e^{-6\alpha\psi}  \left( \psi^{\lambda} F^{\mu\nu}+\psi^{\mu} F^{\lambda\nu} \right)D_{\lambda}F_{\mu\nu}
\nonumber\\&&\quad\quad
+10\alpha e^{-6\alpha\psi}  F^{\mu\lambda}F^{\nu}\,_{\lambda}D_{\mu}\psi_{\nu}
+24\alpha^{2}D_{\mu}\psi_{\nu}D^{\mu}\psi^{\nu}
+4\alpha e^{-6\alpha\psi} \psi^{\mu} F_{\mu\nu}D_{\lambda}F^{\lambda\nu}
\nonumber\\&&\quad\quad
+8\alpha(\alpha\psi_{\mu}\psi_{\nu}-D_{\mu}\psi_{\nu})R^{\mu\nu}
-4\alpha^{2}\psi_{\mu}\psi^{\mu}R
+4\alpha^{2}D_{\mu}\psi^{\mu}D_{\nu}\psi^{\nu}
+48\alpha^{3}\psi_{\mu}\psi^{\mu} D_{\nu}\psi^{\nu}
\nonumber\\&&\quad\quad
-144\alpha^{3}\psi^{\mu}\psi^{\nu} D_{\mu}\psi_{\nu}
+204\alpha^{4}\psi_{\mu}\psi^{\mu}\psi_{\nu}\psi^{\nu}
\Big].
\end{eqnarray}
Finally, we can say that the quadratic term $R_{\mu\nu\lambda\sigma}R^{\mu\nu\lambda\sigma}$ does not have its canonical form in this case, as expected. However, to avoid this problem, we can set $\psi(x)=0$  which is not an acceptable condition in the KK theories,
so that a nonphysical constraint, $F_{\rho\tau}F^{\rho\tau}=0$, arises from the last equation of (\ref{son1}). In fact, this is the well-known inconsistency problem of the KK theories. Conversely, in the considered model, $\mathcal{\widetilde{Q}}_{\nu}$ and $\mathcal{\widetilde{U}}$ in (\ref{son1}) are
not necessarily zero, and even if $\psi(x)=0$, we do not have $F_{\rho\tau}F^{\rho\tau}=0$, but the Lorentz force density term, $\textit{f}_{\lambda}=F_{\lambda}\,^{\rho}{D}_{\tau}F^{\tau}\,_{\rho}$, appears in the last equation of (\ref{lastfieldson}) as
\begin{eqnarray}\label{}
\textit{f}_{\lambda}=-D_{\lambda}(F_{\rho\tau}F^{\rho\tau}),
\end{eqnarray}
which can also be obtained from  equations (\ref{ricciten}) and (\ref{proper3}) if we recall that $\varphi(x)=1$ in\cite{halil13}.

\subsection{The $\beta=0$ case}

Let us investigate the reduced equations by assuming that $\beta=0$. This strong restriction actually breaks  the high-dimensional structure of the metric (\ref{kkmetric3}). Nonetheless, this case still deserves attention from the viewpoint of the considered model, and  $\alpha$ is a free parameter. In this regard, we have from  equation
 (\ref{generalf}) that
\begin{eqnarray}\label{fe3}
 & &D_{\lambda}\mathcal{\widetilde{P}}_{\nu \sigma}-D_{\sigma}\mathcal{\widetilde{P}}_{\nu \lambda}
+ \frac{1}{2}e^{-\alpha\psi} \left( F_{\nu \sigma}\mathcal{\widetilde{Q}}_{\lambda}-F_{\nu \lambda}\mathcal{\widetilde{Q}}_{\sigma}
+ 2 F_{\lambda\sigma}\mathcal{\widetilde{Q}}_{\nu} \right)
+\alpha\Big[\psi^{\rho}\left(g_{\nu\lambda}\mathcal{\widetilde{P}}_{\rho \sigma}
-g_{\nu\sigma}\mathcal{\widetilde{P}}_{\rho \lambda}\right)
\nonumber\\
&&
+\psi_{\lambda}\mathcal{\widetilde{P}}_{\nu \sigma}-\psi_{\sigma}\mathcal{\widetilde{P}}_{\nu \lambda}\Big]=0
,\label{pp1} \nonumber \\
 & & D_{\lambda}\mathcal{\widetilde{Q}}_{\nu}
 +\alpha g_{\nu\lambda}\psi^{\rho} \mathcal{\widetilde{Q}}_{\rho}
 -\alpha\psi_{\nu} \mathcal{\widetilde{Q}}_{\lambda}
  +\frac{1}{2}e^{-\alpha\psi}   F_{\nu}\,^{\rho} \mathcal{\widetilde{P}}_{\rho\lambda}
 -\frac{1}{2}e^{-\alpha\psi} F_{\nu\lambda}\mathcal{\widetilde{U}}=0, \label{pp3}\nonumber  \\
 & & D_{\lambda}\mathcal{\widetilde{U}}
 -\frac{1}{2} e^{-\alpha\psi} F_{\lambda}\,^{\rho}\mathcal{\widetilde{Q}}_{\rho}
=0,\label{pp4}
\end{eqnarray}
where the corresponding equations of motion of the unified theory are equal to
\begin{eqnarray}\label{son2}
&&\mathcal{\widetilde{P}}_{\mu\nu}
=e^{-2\alpha\psi}R_{\mu\nu}
-\frac{1}{2}e^{-4\alpha\psi}
F_{\mu\rho}F_{\nu}\,^{\rho}
- \alpha e^{-2\alpha\psi}\big[  g_{\mu\nu}D_{\rho}\psi^{\rho}
+2 D_{\mu}\psi_{\nu}\big]
\nonumber\\&&\qquad\quad
-2\alpha^{2}e^{-2\alpha\psi}\big[ g_{\mu\nu}\psi_{\rho}\psi^{\rho}
-\psi_{\mu}\psi_{\nu}\big]
,\nonumber\\
&&\mathcal{\widetilde{Q}}_{\nu}
=-\frac{1}{2} e^{-3\alpha\psi}{D}_{\rho}F^{\rho}\,_{\nu}
 ,\nonumber\\
&&\mathcal{\widetilde{U}}
=\frac{1}{4}e^{-4\alpha\psi} F_{\rho\tau}F^{\rho\tau},
\end{eqnarray}
in the four dimensions. Moreover, the final form of the action (\ref{lastaction}) is similarly reorganized as:
\begin{eqnarray}\label{soniva}
&&
\widetilde{\hat I}=\frac{1}{2\hat\kappa^{2}}\int_{\hat{\mathcal{M}}}d^{4}x\,dy\,\sqrt{-{ g}}
\,\Big[R_{\mu\nu\lambda\sigma}R^{\mu\nu\lambda\sigma}
-\frac{3}{2}e^{-2\alpha\psi} R_{\mu\nu\lambda\sigma}F^{\mu\nu}F^{\lambda\sigma}
+\frac{3}{8}e^{-4\alpha\psi}F_{\mu\nu}F^{\mu\nu}F_{\lambda\sigma}F^{\lambda\sigma}
\nonumber\\&&\quad\quad
+\frac{5}{8}e^{-4\alpha\psi}F_{\mu\nu}F^{\nu\lambda}F_{\lambda\sigma}F^{\sigma\mu}
+3\alpha^{2}e^{-2\alpha\psi}\left(3\psi_{\lambda}\psi^{\lambda} F_{\mu\nu}F^{\mu\nu}
+2\psi_{\mu}\psi^{\lambda}F_{\lambda\nu}F^{\nu\mu}\right)
\nonumber\\&&\quad\quad
+e^{-2\alpha\psi}D_{\lambda}F_{\mu\nu}D^{\lambda}F^{\mu\nu}
- 4\alpha e^{-2\alpha\psi} \left( \psi^{\lambda} F^{\mu\nu}+\psi^{\mu} F^{\lambda\nu} \right)D_{\lambda}F_{\mu\nu}
+ 6\alpha e^{-2\alpha\psi} F^{\mu\lambda}F^{\nu}\,_{\lambda}D_{\mu}\psi_{\nu}
\nonumber\\&&\quad\quad
+4\alpha^{2}\left(2D_{\mu}\psi_{\nu}D^{\mu}\psi^{\nu}
+D_{\mu}\psi^{\mu}D_{\nu}\psi^{\nu}\right)
+4\alpha e^{-2\alpha\psi}\psi^{\mu} F_{\mu\nu}D_{\lambda}F^{\lambda\nu}
+8\alpha(\alpha\psi_{\mu}\psi_{\nu}-D_{\mu}\psi_{\nu})R^{\mu\nu}
\nonumber\\&&\quad\quad
-4\alpha^{2}\psi_{\mu}\psi^{\mu}R
+16\alpha^{3}\left(\psi_{\mu}\psi^{\mu} D_{\nu}\psi^{\nu}
-\psi^{\mu}\psi^{\nu} D_{\mu}\psi_{\nu}\right)
+12\alpha^{4}\psi_{\mu}\psi^{\mu}\psi_{\nu}\psi^{\nu}
\Big].
\end{eqnarray}
In this case, the coefficient of the gravitational part $R$, which does not have its canonical form, becomes  $e^{2\alpha\psi}$ in (\ref{act3}), but the $R_{\mu\nu\lambda\sigma}R^{\mu\nu\lambda\sigma}$ term is canonical in (\ref{soniva}). The equation (\ref{son2}) also gives  the ${D}_{\rho}F^{\rho}\,_{\nu}=0$ and $F_{\rho\tau}F^{\rho\tau}=0$, if $\mathcal{\widetilde{Q}}=0$ and $\mathcal{\widetilde{U}}=0$, respectively,  despite $\psi(x)\neq0$.
As mentioned before, this is not a problem for the WYKK theory due to the structure of general field equations (\ref{fe3}). Actually,  $\mathcal{\widetilde{U}}\neq0$ is the necessary condition in this case; otherwise  all equations in (\ref{fe3}) fall into triviality. Besides, we can now determine the following new Lorentz force density term including boson fields as follows:
\begin{eqnarray}\label{}
\textit{f}_{\lambda}=-D_{\lambda}(\psi F_{\rho\tau}F^{\rho\tau}),
\end{eqnarray}
from  last equations of (\ref{fe3}) and (\ref{son2}) if we choose the parameter $\alpha=-1/4$, and without considering, certainly, $\psi(x)\neq0$.

\section{Conclusion}\label{sec6}
In this work, by employing an alternative form of the basis equation of the considered model in the review section, we have recomputed the reduced field equations in terms of the new $\left\{\mathcal{\overline{P}}_{\mu\nu},\mathcal{\overline{Q}}_{\nu},\mathcal{\overline{U}}\right\}$ set, i.e., the equations of the typical KK theory. The new expressions (\ref{proper1})$–$(\ref{proper3}) seem to be more accurate than others, including the $\left\{\mathcal{{P}}_{\mu\nu},\mathcal{{Q}}_{\nu},\mathcal{{U}}\right\}$ set in \cite{halil13}. Then, using simplifications that result from the horizontal lift basis, we have generalized the resulting equations to the Einstein frame in which the conformal metric ansatz contains arbitrary parameters $\alpha$  and $\beta$. After these reduction procedures, we have derived the desired set of field equations, which are governed by the new conformal $\{\mathcal{\widetilde{P}}_{\mu\nu},\mathcal{\widetilde{Q}}_{\nu},
\mathcal{\widetilde{U}}\}$ set, and the transformed quadratic action in the coordinate basis both by considering the vielbeins fields and the conformal transformation rules. Conversely, we have investigated the consequences of the two possible cases, $\beta=-2\alpha$ and $\beta=0$, on this approach. As we mentioned in our previous work \cite{halil13}, for the condition $\psi(x)=0$, the standard KK theory renders the well-known nonphysical constraint, $F_{\rho\tau}F^{\rho\tau}=0$, whereas this Lorentz invariant term does not have to be equal to zero in the WYKK theory because of the more general field equation (\ref{lastfieldson}). By contrast, the Lorentz force density, $\textit{f}_{\lambda}=F_{\lambda}\,^{\rho}{D}_{\tau}F^{\tau}\,_{\rho}$, appears for the former case naturally. Additionally, even if $\psi(x)\neq0$, we have demonstrated that the density term can only be equal to the negative gradient of the new invariant, i.e., $\textit{f}_{\lambda}=-D_{\lambda}(\psi F_{\rho\tau}F^{\rho\tau})$ for the latter case with $\alpha=-1/4$.

Let us finally remark that we can extend our formalism to investigate the dimensionally reduced form of equation (\ref{fe4e}), whose second term had already been calculated in (\ref{inv2}), in the 5D generalized KK theory for completeness. Furthermore, whether the first equation of (\ref{generalf}) can be written in the form of the equation (\ref{fe5e}) can be studied. We have already shown that \cite{kuyrukcu16} the first field equations, i.e.,
\begin{eqnarray}
 D_{\lambda}\mathcal{{P}}_{\nu \sigma}-D_{\sigma}\mathcal{{P}}_{\nu \lambda}
 -\frac{1}{4}\varphi^{2} \left( F_{\nu \sigma}\mathcal{{Q}}_{\lambda}-F_{\nu \lambda}\mathcal{{Q}}_{\sigma}
+ 2 F_{\lambda\sigma}\mathcal{{Q}}_{\nu} \right)=0,
\end{eqnarray}
which is given by Başkal and Kuyrukcu\cite{halil13} or can be obtained from (\ref{proper1}) using $\mathcal{\overline{P}}_{\mu\nu}=\mathcal{P}_{\mu\nu}$ and $\mathcal{\overline{Q}}_{\nu}=-(1/2)\varphi\mathcal{Q}_{\nu}$, can exactly be written in the following Camenzind’s current density form
\begin{eqnarray}\label{}
D_{\mu}R^{\mu}\,_{\nu\lambda\sigma}=D_{\lambda}\left(T_{\nu\sigma}-\frac{1}{2}g_{\nu\sigma}T\right)
-D_{\sigma}\left(T_{\nu\lambda}-\frac{1}{2}g_{\nu\lambda}T\right),
\end{eqnarray}
for the case in which $\mathcal{{Q}}_{\lambda}=0$, $\mathcal{{U}}=0$ or more generally $\mathcal{{Q}}_{\lambda}=0$, and $D_{\lambda}\mathcal{{U}}=0$. Here $T_{\nu\sigma}$ represents the stress-energy tensor that comes from the KK theory of gravity \cite{Liu:1997fg}. We are working on obtaining such relationships for the model under consideration.

\section{Acknowledgments}

The author would like to thank the anonymous referees for the constructive comments.

\end{document}